\documentclass[journal=mamobx, layout=twocolumn, manuscript=article]{achemso}
\pdfoutput=1
\usepackage[utf8]{inputenc}
\usepackage[T1]{fontenc}
\usepackage{color}
\usepackage{graphicx}
\usepackage{amsmath, amssymb}
\usepackage{ulem}
\usepackage[dvipsnames]{xcolor}

\usepackage{xr-hyper}

\makeatletter

\newcommand*{\addFileDependency}[1]{
\typeout{(#1)}
\@addtofilelist{#1}
\IfFileExists{#1}{}{\typeout{No file #1.}}
}\makeatother

\newcommand*{\myexternaldocument}[1]{%
\externaldocument{#1}%
\addFileDependency{#1.tex}%
\addFileDependency{#1.aux}%
}

\myexternaldocument{zSI}

\setkeys{acs}{doi = false}
\usepackage{cuted}
\usepackage{placeins}
\usepackage[version=4]{mhchem}
\usepackage{cleveref}
\usepackage{mathtools}
\usepackage[sort&compress,numbers,super]{natbib}
\SectionNumbersOn

\title{2D capsid formation within an oscillatory energy landscape: orderly self-assembly depends on the interplay between a dynamic potential and intrinsic relaxation times.}

\author{Jessica K. Niblo}
\affiliation[University of Virginia]
{Department of Chemistry, University of Virginia, Charlottesville, VA}
\author{Jacob R. Swartley}
\affiliation[University of Virginia]
{Department of Chemistry, University of Virginia, Charlottesville, VA}
\author{Zhongmin Zhang}
\affiliation[UNC]
{Department of Chemistry, University of North Carolina at Chapel Hill, Chapel Hill, NC}
\author{Kateri H. DuBay}
\email{dubay@virginia.edu}
\affiliation[University of Virginia]
{Department of Chemistry, University of Virginia, Charlottesville, VA}

\begin{document}

\maketitle

\begin{abstract}

Multiple dissipative self-assembly protocols designed to create novel structures or to reduce kinetic traps have recently emerged.
Specifically, temporal oscillations of particle interactions have been shown effective at both aims, but investigations thus far have focused on systems of simple colloids or their binary mixtures. In this work, we expand our understanding of the effect of temporally oscillating interactions to a two-dimensional coarse-grained viral capsid-like model that undergoes a self-limited assembly. This model includes multiple intrinsic relaxation times due to the internal structure of the capsid subunits and, under certain interaction regimes, proceeds via a two-step nucleation mechanism.  
We find that oscillations much faster than the local intrinsic relaxation times can be described via a time averaged inter-particle potential across a wide range of interaction strengths, while oscillations much slower than these relaxation times result in structures that adapt to the attraction strength of the current half-cycle. 
Interestingly, oscillation periods similar to these relaxation times shift the interaction window over which orderly assembly occurs by enabling error correction during the half-cycles with weaker attractions. Our results provide fundamental insights to non-equilibrium self-assembly on temporally variant energy landscapes.

\end{abstract}

\section{Introduction}\label{sec:intro}

Self-assembly is the process by which a disordered system forms ordered patterns or nanostructures without external intervention due to the interactions that are encoded within the assembling components and their environment. The driving forces behind self-assembly can organize lipids into bilayers,\cite{cho2013a} 
gather capsomers into viral capsids,\cite{zlotnick1994,zlotnick2000,hicks2006, hagan2006, sigl2021} and arrange block copolymers into a wide range of microphase topologies.\cite{lynd2008} 

Several studies have worked to uncover governing principles that would enable the design of interparticle interactions that lead to well-ordered, self-assembled equilibrium states. 
As a result, we now know that the strength, placement, and specificity of the interactions between the assembling components,\cite{zhang2004,glotzer2007,hormoz2011,haxton2012,grunwald2014,mallory2016}, their shapes,\cite{zhang2004, glotzer2007,damasceno2012, grunwald2014} and their concentrations \cite{hagan2006, hagan2010} can all be tuned to stabilize a specific equilibrium target structure. 
However, the ability to reach these equilibrium assemblies 
is highly dependent upon the assembly kinetics.\cite{hagan2006,grant2011} Strengthening the interactions that lower the free energy of the target structure often increases the kinetic barriers to its formation, 
 making it a challenge to design components that reliably self-assemble on a reasonable timescale.\cite{pawar2010,hormoz2011,grant2011,whitelam2015a}
Insightful work has been done to characterize the different dynamic pathways a system can take during self-assembly and the various types of kinetic traps that may emerge.\cite{whitelam2015a,hagan2011} 
One recent article has even been able to optimize interactions that not only select a target equilibrium structure, but also simultaneously control specific kinetic features along its assembly pathway.\cite{goodrich2021}

Alternatively, dissipative self-assembly processes can result in well-ordered structures by driving a system out of equilibrium and either creating new assembly routes towards equilibrium structures or forming non-equilibrium steady-states (NESSs).\cite{whitesides2002a,fialkowski2006}   
These non-equilibrium self-assembly pathways couple an assembling system to an energy source, such as when particles are self-propelled with a constant directional force\cite{mallory2016,ilse2016,lowen2018} or when assembly occurs within a shear flow.\cite{das2021b, das2023} Recently, large deviation theory has been successfully employed to optimize both interactions and external shear forces in order to target specific steady states.\cite{das2021b}

The energy source in dissipative self-assembly may be temporally modulated through the change of an internal or external parameter, such as environmental changes that modify the interactions between assembling particles\cite{tagliazucchi2014,tagliazucchi2016,long2018} or that change their interactions with an external field.\cite{promislow1996, promislow1997,swan2012,swan2014,bauer2015, kim2020}
The self-assembly of many biological structures occurs within complex and ever-changing environments, and even certain naturally-forming non-biological materials appear to require time-variant environments to assemble -- recent investigations have revealed that cycling between undersaturation and supersaturation 
may be necessary for the formation of naturally-occurring dolomite.\cite{garcia-ruiz2023,kim2023}

One way to optimize a time-dependent dissipative self-assembly pathway is to employ feedback control, in which the assembly process is monitored in real time and the driving forces are adjusted on the fly, based on that feedback, to guide the process towards a desired outcome. 
One such approach adjusted the inter-particle interactions during assembly simulations based on the ratio of correlation and response functions and the degree to which they indicated an optimal balance of local microscopic reversibility (to avoid kinetic traps) and overall global irreversibility (towards assembly).\cite{klotsa2013}  
More generally, it has been shown that high dimensional non-equilibrium time-dependent forces are able to guide an assembly towards a desired structure, however there is an unavoidable energetic cost to doing so.\cite{chennakesavalu2021} 
Bevan and coworkers constructed an experimental system to demonstrate the feasibility of feedback control: the 2D assembly of charged colloids were monitored in real time via optical microscopy while being subjected to a tunable electric potential. Within this set-up, the electric potential was adjusted based on feedback from the structural order parameter in order to construct perfectly crystalline configurations.\cite{juarez2012, xue2014, tang2016a} 
Even so, it is clear that the need to monitor and adjust assembly conditions in real time would present significant difficulties to widely implementing feedback control approaches.

As an alternative to monitoring and adjusting each assembly process in real-time, information can be gathered from multiple simulations or experimental realizations of a dissipative assembly process and used to construct time-dependent protocols that are optimized for the ensemble of likely assembly pathways. One such approach employed evolutionary reinforcement learning to determine time-dependent temperature and chemical potential protocols for the efficient assembly of patchy disks into desired polymorphs.\cite{whitelam2020}
Similarly, Markov state models have been constructed from simulations of colloidal oligomers and capsids in order to design optimal time-dependent interaction profiles for the finite-time folding and assembly of those systems.\cite{trubiano2022} However, significant data is required in order to optimize each time-dependent protocol, which will depend on the specifics of the system, and the implementation of these time-dependent protocols in real systems may prove challenging.

Oscillatory or cyclic changes to the 
environment in which assembly proceeds may provide a more experimentally accessible avenue to design new dissipative assembly pathways.
A number of studies have shown that the cyclical exposure of oil droplets to an external magnetic field can facilitate local relaxation processes in the resulting aggregates, thereby enabling them to overcome kinetic barriers to equilibration.\cite{promislow1996,promislow1997,swan2012, swan2014, bauer2015, kim2020,martin-roca2023} 
In simulations of photosensitive nanoparticles, light-induced aggregation proceeded more rapidly after short pauses in the light irradiation.\cite{jha2012} 
Simulations by Risbud and Swan found that toggling inter-particle depletion interactions on and off at a timescale that allowed for sufficient particle diffusion relaxed kinetic traps and resulted in the more rapid formation of low-defect colloidal crystals.\cite{risbud2015} 
Finally, joint experiment and simulation work showed that the cyclic toggling of an external electric field on a timescale close to the characteristic melting time could anneal defects in colloidal crystals.\cite{kao2021}

Temporal environmental oscillations may also provide experimentally accessible ways to create and maintain long-lived non-equilibrium steady states (NESSs).
Tagliazucchi and coworkers used simulations and theory to show that novel non-equilibrium steady state phases of binary pH-sensitive colloids can form when pH oscillations are faster than the colloid's characteristic diffusional timescale,\cite{tagliazucchi2014, tagliazucchi2016} and a similar effect was observed with three dimensional close packed colloids.\cite{long2018}
Additional simulations have shown that oscillations in inter-particle interactions and external fields can result in ellipsoids adopting a non-equilibrium chiral smectic phase\cite{chen2022a} and in the formation of non-equilibrium lamellar structures in homopolymer mixtures.\cite{pigard2019a}

Prior work has focused on the assembly of extended NESS structures or the annealing of defects in extended colloidal crystals. In this work, however, we investigate the assembly of a self-limited capsid-like model with multiple inherent length scales and relaxation timescales that arise from the internal structure of the capsid building blocks.  
Specifically, we probe the impact of oscillatory time-dependent interactions on the assembly mechanisms over a range of oscillation timescales and amplitudes. 
In Section~\ref{sec:methods}, we outline the coarse grained viral capsid-like model.  
In Section~\ref{sec:StaticSystem}, we discuss assembly with static interactions. 
Then, in Section~\ref{sec:oscperiod}, we perform simulations with oscillatory interactions that are faster than, slower than, or similar to inherent relaxation timescales within the model to probe their influence on assembly.  
Finally, in Section~\ref{sec:amplitude}, we vary oscillation amplitude to investigate how these temporal oscillations can act as an error correction technique. 
As summarized in Section~\ref{sec:conc}, we find that the inherent timescales of the model system are critical in determining the effect of fast, intermediate, and slow oscillatory interactions on the resulting assembly process.

\section{Modeling} \label{sec:methods}

\begin{figure}[h]
\centering
  \includegraphics[height=8cm]{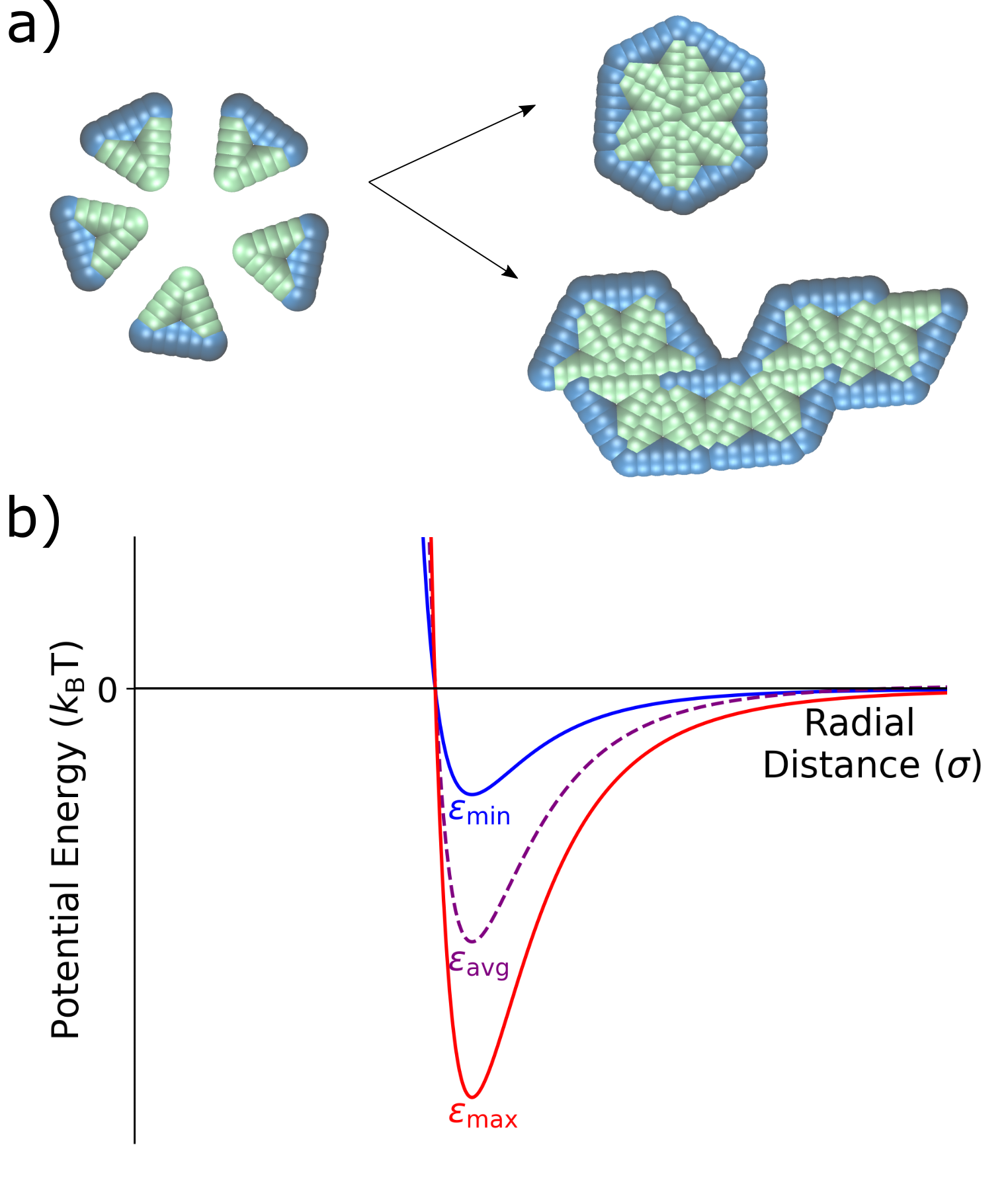}
  \caption{{\bf Schematic of the coarse-grained capsid-like model and inter-particle potential.} 
  (a) Triangular particles with two type A edges (green) 
  and one type B edge (blue) 
  start in a random distribution within the simulation box. Type A - type A interactions are defined via a Lennard-Jones potential, while the type A - type B and type B - type B particles interact via a Weeks-Chandler-Andersen potential.\cite{weeks1971} As the simulation proceeds, triangles can assemble into hexameric capsid-like structures or into larger snake-like aggregates. 
  (b) The Lennard-Jones potential for the type A - type A interaction, showing the two different values of $\epsilon$ -- $\epsilon_{\mathrm{min}}$ and $\epsilon_{\mathrm{max}}$ -- with the time-averaged $\epsilon_{\mathrm{avg}}$ shown in purple.  
  }
  \label{fgr:structures}
\end{figure}

Viral capsid models have proven essential in the theoretical and computational study of self-assembly, due in part to their organization into self-limiting ordered structures.\cite{hagan2006,jack2007,nguyen2007,rapaport2004,rapaport2008,hagan2010,krishna2010a,hagan2011,perlmutter2015,pak2019,hagan2021}  
Inspired by earlier studies on capsid assembly, in this work we investigate the assembly of two-dimensional, rigid triangular particles that assemble into capsid-like hexamers, as shown in Fig.~\ref{fgr:structures}a. The model was based upon a similar one in Mallory and Cacciuto's work on the role of self-propulsion in capsid-like colloidal assembly.\cite{mallory2016} 
The placement of attractive particles (green) on only two edges of the triangles causes these monomer units to assemble either into distinct hexamers or into extended snake-like structures (see Fig.~\ref{fgr:structures}a).

The triangular monomers are composed of fifteen partially overlapping circular subparticles that are rigidly held together, as shown in Fig.~\ref{fgr:structures}a. Each subparticle is assigned one of two different types, A or B, which determines its interparticle potential. 
The subparticles on two edges of the equilateral triangle are assigned type A (green) and interact via a Lennard-Jones (LJ) potential, the strength of which can be tuned by changing $\epsilon$, the LJ well-depth. 
The subparticles on the third edge are assigned type B (blue) and interact via the Weeks-Chandler-Andersen (WCA) potential \cite{weeks1971}. Thus, type B subparticles are purely repulsive, and interactions between type A and type B subparticles are also defined by the WCA potential. 
The placement of these differently-interacting subparticles introduces an overall anisotropic interaction between the individual triangular particles. 
Fig.~\ref{fgr:structures}b illustrates the attractive interactions between the type A subparticles of different triangular particles for various $\epsilon$ values. A more detailed descriptions of the particles and interactions can be found in the Supporting Information.

Oscillatory interactions are implemented by switching the $\epsilon$ value of the type A-type A interactions between $\epsilon_{\mathrm{min}}$ and $\epsilon_{\mathrm{max}}$ in a square wave pattern. We investigate the effect on capsid formation of variations in both the oscillation period and its amplitude (defined as the magnitude of the shift in each direction from the central, $\epsilon_{\mathrm{avg}}$-value).  
Variations in $\epsilon$ are considered here as a way to investigate the effects of a temporally-dependent interaction strength on assembly.

Simulations begin with a randomly distributed system of 150 triangular particles in a periodic box with no particle overlap. Langevin Dynamics is utilized to evolve the system in time, which also 
provides a thermostat to maintain a constant temperature over the simulation and mimics the drag and random fluctuations associated with the dynamics of solvated particles. The system is first allowed to equilibrate with the type A interactions turned off and all subparticles interacting via the WCA potential, 
as this ensures that there is no particle overlap while establishing a random initial distribution of the triangles, both spatially and orientationally. After the equilibration period, the Type A attractive interactions are turned on, and the system is progressed for 150,000$\tau$, where $\tau$ is time in reduced units.
See Supporting Information for additional details.

\section{Results and Discussion}

\subsection{Capsid assembly is non-monotonic with attraction strength.} \label{sec:StaticSystem}

\begin{figure}[h]
\centering
  \includegraphics[height=6.9cm]{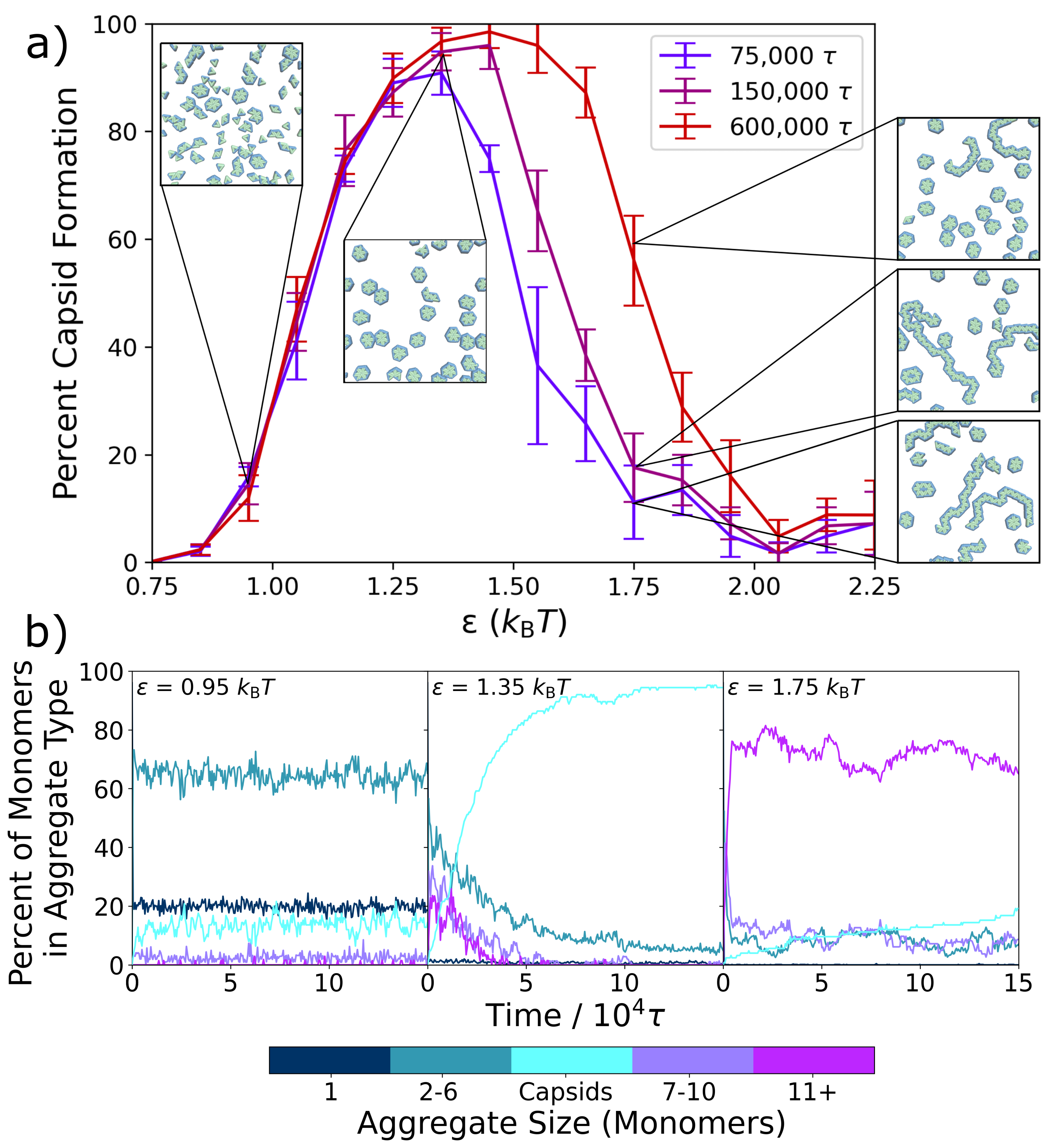} 
  \caption{{\bf Capsid yield and aggregate formation within the static system.} a) Capsid yield curves are plotted after three different simulation times vs. the attraction strength, $\epsilon$. As simulation time increases, the capsid yield does not change on the left hand side of the curve. However, on the right hand side, capsid yield increases with longer simulation time. Inset are final snapshots of the system at $\epsilon=0.95k_{\mathrm{B}}T$, $\epsilon=1.35k_{\mathrm{B}}T$, and $\epsilon=1.75k_{\mathrm{B}}T$. b) The percentage of triangular monomers within each group of different sized aggregates is plotted over 150,000$\tau$ for three different $\epsilon$ values ($\epsilon=0.95k_{\mathrm{B}}T$, $\epsilon=1.35k_{\mathrm{B}}T$, and $\epsilon=1.75k_{\mathrm{B}}T$). Aggregates of varying size are shown with the color bar, with the hexameric structure shown in cyan. Capsid yields and kinetic traces are averaged over five independent trajectories, and the error bars in (a) display the standard deviation.}
  \label{img:equilYield}
\end{figure}

We first model the formation of the complete capsid-like hexamers within a series of static environments, each with fixed effective interactions, ranging from $\epsilon=0.75k_{\mathrm{B}}T$ to $\epsilon=2.25k_{\mathrm{B}}T$. 
Figure~\ref{img:equilYield}a shows the resulting capsid yields and sample configurations as a function of $\epsilon$ for three different simulation times, and Figure~\ref{img:equilYield}b shows how the presence of differently-sized aggregates changes with $\epsilon$ and simulation time.

In keeping with prior work on similar models of viral capsids\cite{mallory2016,hagan2006,rapaport2008,nguyen2007,hagan2011,perlmutter2015}, we observe in Figure~\ref{img:equilYield}a that capsid assembly is non-monotonic with interaction strength. 
As $\epsilon$ increases from $0.75k_{\mathrm{B}}T$ to $1.35k_{\mathrm{B}}T$, capsid yield also increases. However, above $\epsilon = 1.35k_{\mathrm{B}}T$, capsid yield decreases. This non-monotonic trend is due to a transition from thermodynamically equilibrated systems on the left-hand side of the curve to kinetically constrained systems on the right-hand side as interactions become too strong for full equilibration to occur within the simulation time. Sampling for various lengths of time verifies this transition from thermodynamic to kinetic products, as the left-hand side of the curve remains unchanged for sampling times ranging from 75,000$\tau$ to 600,000$\tau$, while the right-hand side shifts to higher values as sampling times lengthen and the kinetically constrained structures have additional time to relax. 

The inset snapshots in Figure~\ref{img:equilYield}a provide further evidence of this transition. 
At low $\epsilon$ interaction strengths, such as shown for $\epsilon = 0.95k_{\mathrm{B}}T$, most triangles exist as free monomers or in small aggregates, and only a few complete capsids form. 
By $\epsilon = 1.35k_{\mathrm{B}}T$, the snapshot shows almost complete capsid formation. However, at $\epsilon = 1.75k_{\mathrm{B}}T$, well-formed capsids appear alongside larger, snake-like and kinetically trapped aggregates, with the number of snake-like aggregates decreasing and the number of capsids increasing as simulation times lengthen.

To better probe capsid assembly kinetics, in Figure~\ref{img:equilYield}b we track the percent of the triangular monomers that are assembled into differently sized aggregates as a function of simulation time. At a weaker interaction strength of $\epsilon=0.95 k_{\mathrm{B}}T$, the system almost instantaneously assembles into small aggregates of 2-6 monomers (turquoise) in length, which slightly decreases over the first $1\times10^4\tau$ of the simulation as a small number of capsids (cyan) assemble. At an intermediate attraction strength of $\epsilon = 1.35k_{\mathrm{B}}T$, a significant number of longer aggregates composed of 7-10 (purple) and 11+ (magenta) monomers rapidly form initially, but then decrease on the same timescale as capsid structures emerge. At a stronger attraction of $\epsilon = 1.75k_{\mathrm{B}}T$, where the capsid yield curve has crossed into the kinetic regime, most monomers rapidly assemble into the largest aggregates (magenta), which only slightly and gradually convert into capsids over the remainder of the simulation time. By $150,000\tau$, most monomers remain in these very large aggregates due to the difficulties in overcoming the high energetic barriers associated with strong inter-particle interactions.

Interestingly, at the intermediate and stronger attraction values, the assembly process in Figure~\ref{img:equilYield}b follows a non-classical, two-step assembly pathway.\cite{kashchiev2005, deyoreo2015} During non-classical assembly pathways, which have been seen in protein \cite{haxton2012, dogan2014, wilson2018} and colloidal particle crystallization,\cite{sherman2016, sherman2019, savage2009} the system assembles into a condensed, disordered phase that has a lower free energy barrier to nucleation than the disassembled system, so that the final assembled structure grows from pre-formed aggregates.\cite{sherman2019, savage2009, wilson2018} This process is clearly in evidence at both $\epsilon = 1.35k_{\mathrm{B}}T$ and $\epsilon = 1.75k_{\mathrm{B}}T$ in Figure~\ref{img:equilYield}b, where capsid counts increase as the initially assembled larger aggregates decrease. To probe exactly the $\epsilon$ values at which this changeover in assembly mechanism occurs, we expand Figure~\ref{img:equilYield}b in Figure~\ref{img:EquilForm} and plot the results for values ranging from $\epsilon = 0.75k_{\mathrm{B}}T$ to $\epsilon = 2.15k_{\mathrm{B}}T$ at $0.10k_{\mathrm{B}}T$ intervals. Evidence for this two-step mechanism can be seen as early as $\epsilon = 1.25k_{\mathrm{B}}T$, and a clear cross-over is seen around $\epsilon = 1.45k_{\mathrm{B}}T$, where non-capsid monomers are present in smaller and larger aggregates in about equal numbers, with both types of aggregates decreasing as capsids form. At higher $\epsilon$ values, the two-step pathway dominates, with most monomers assembling rapidly into the larger, snake-like aggregates.

Having established the assembly behavior of the 2D model capsid system within a series of static environments, we now temporally vary the inter-monomer attractions by switching the strength of the Lennard-Jones potential between two values during assembly. 
We define the stronger attraction strength as $\epsilon_{\mathrm{max}}$ and the weaker interaction strength as $\epsilon_{\mathrm{min}}$. The oscillation amplitude defines the distance of $\epsilon_{\mathrm{max}}$ and $\epsilon_{\mathrm{min}}$ from a central $\epsilon_{\mathrm{avg}}$ value, where $\epsilon_{\mathrm{avg}} = (\epsilon_{\mathrm{max}}+\epsilon_{\mathrm{min}})/2$. See Figure~\ref{fgr:structures}. We also define the period of oscillation, $\tau_{\mathrm{osc}}$, as the time it takes to complete a full cycle, with half the cycle at $\epsilon_{\mathrm{max}}$ and the other half at $\epsilon_{\mathrm{min}}$.
In the rest of the paper, we investigate how these oscillatory interactions influence capsid formation across a variety of oscillation periods (Sections~\ref{sec:oscperiod}) and oscillation amplitudes (Section~\ref{sec:amplitude}).

\subsection{Assembly depends on the oscillation frequency.} \label{sec:oscperiod}

First, we consider how different oscillation periods, $\tau_{\mathrm{osc}}$, affect the assembly process.
During the $\epsilon_{\mathrm{max}}$ half-cycle, triangular monomers are more strongly attractive and assemble together, while in the $\epsilon_{\mathrm{min}}$ half-cycle, such structures may rearrange or be broken apart. The degree to which assembly and disassembly occurs depends upon the time spent in each half-cycle, and how that time compares to the time required for the system to equilibrate.

Previous studies investigating oscillations in inter-colloidal potentials compare $\tau_{\mathrm{osc}}$ to a characteristic diffusional time-scale, $t_{\mathrm{d}} = {\sigma^2}/{D}$, where $\sigma$ specifies the size of  
the assembling colloids and $D$ is the diffusion coefficient.\cite{tagliazucchi2014,risbud2015} 
Oscillation periods that are significantly shorter than the characteristic diffusional time-scale ($\tau_{\mathrm{osc}}<<t_{\mathrm{d}}$) are considered to be at the fast oscillation limit, while at the slow oscillation limit, periods are long enough to allow the system to relax to different equilibrium structures during each oscillation half-cycle.\cite{tagliazucchi2014, risbud2015, tagliazucchi2016}

As can be seen in the structure shown in Figure~\ref{fgr:structures}, more than one length-scale is needed to fully describe the assembling particles; the diameter of the circular subparticles is $\sigma_{\mathrm{LJ}}=$0.25$\sigma$, while the edge length of the triangular particles is 1$\sigma$. 
In addition, at the higher attraction strengths we probe, capsid formation proceeds via the initial formation of a condensed disordered phase (see Fig.~\ref{img:equilYield}a) and its subsequent relaxation.  
The multiple length-scales and assembly pathways within this model are expected to influence how capsid formation varies with oscillation frequency.

To better understand the interplay of the oscillation period and these inherent length-scales, we list in Table~\ref{tbl:diffusiontable} a set of key distances that characterize important energetic and structural changes, ranging from the distance required to reduce the attractive interaction to a tenth of its maximum strength to the full edge length of a single triangular particle. 
To estimate how long it would take for the triangular particles to traverse these distances, we plot the mean squared displacement vs. time for a single triangular particle in Figure~\ref{img:MSD}a. 
Since the relevant length scales span the ballistic and diffusive time regimes, we estimate the time it takes for the particle to move over these critical distances directly from the results in Figure~\ref{img:MSD}a. 
Additionally, we plot in Figure~\ref{img:MSD}b the rotational autocorrelation function vs. time for a single triangular particle and calculate the mean rotational lifetime using an exponential decay fit.
Below, we first consider oscillations at the fast limit, where the period is shorter than the timescales associated with these key distances ($\tau_{\mathrm{osc}} << \{t_{\mathrm{d}}\}$), and show that the assembly behavior in this regime can be described by the $\epsilon$ value averaged over a single oscillation period, $\epsilon_{\mathrm{avg}}$. Next, we investigate slow oscillations, where the period is longer than the timescales that correspond to the distances in Table~\ref{tbl:diffusiontable} ($\tau_{\mathrm{osc}} >> \{t_{\mathrm{d}}\}$), such that the system has time to at least partially equilibrate to the current $\epsilon$ value during each half period of the oscillation.
Lastly, we simulate oscillations in the intermediate regime ($\tau_{\mathrm{osc}} \approx \{t_{\mathrm{d}}\}$) and show how the window of capsid assembly shifts with the oscillation period.

\begin{table}[h]
\small
  \caption{\textbf{Characteristic lengths and their corresponding relaxation times.} 
Important length-scales within the model 
are provided along with the estimated time, $t_{\mathrm{d}}$, that it takes for a single, isolated, triangle particle  
to move over that distance or to lose its rotational orientation, based on its mean squared displacement and its rotational autocorrelation function (see Figure~\ref{img:MSD}). 
    }
  \label{tbl:diffusiontable}
  \begin{tabular*}{0.48\textwidth}{@{\extracolsep{\fill}}lll}
    \hline
    Characteristic Distance & Distance ($\sigma$) & $t_{\mathrm{d}}$ $(\tau)$ \\
    \hline
     0.72$\sigma_{\mathrm{LJ}}$ ($-\epsilon \rightarrow -0.1\epsilon$) & 0.18 & 0.59\\
    Diameter of Subparticle & 0.25 & 0.91 \\
 
    Edge of Triangle & 1.00 & 10.33 \\ 
    \hline
    Rotational Relaxation\footnote[3] & -- & 7.10\\
    \hline
  \end{tabular*}
  
 \footnote[3]{}{\footnotesize For the rotational relaxation, $t_{\mathrm{d}}$ describes the mean lifetime for the exponential decay from the rotational correlation function in Figure~\ref{img:MSD}b.
 }
\end{table}

{\bf The fast oscillation limit obtains the equilibrium yield curve for a well-depth of $\epsilon_{\mathrm{avg}}$.}  
In prior work on colloidal assembly under an oscillatory potential, Szleifer and coworkers employed
the Fokker-Planck equation while dividing up the oscillation period into a series of infinitesimally small time steps to show that, when the period of oscillation is much, much shorter than the inherent diffusional time-scale of the simulated colloidal particles, the oscillatory interaction potential can be described by a static effective inter-particle potential that is equal to the time-averaged potential over a single oscillation period.\cite{tagliazucchi2014,tagliazucchi2016} 
In our system, where we oscillate the well-depth of the Lennard-Jones potential such that half of each oscillation period is at a strength of $\epsilon_{\mathrm{max}}$ and half at a strength of $\epsilon_{\mathrm{min}}$, 
the time-averaged potential over a single period works out to be the LJ potential with a well depth of $\epsilon_{\mathrm{avg}} = (\epsilon_{\mathrm{max}} + \epsilon_{\mathrm{min}})/2$  
(see SI for details).

\begin{figure}[h!]
\centering
  \includegraphics[height=6.85cm]{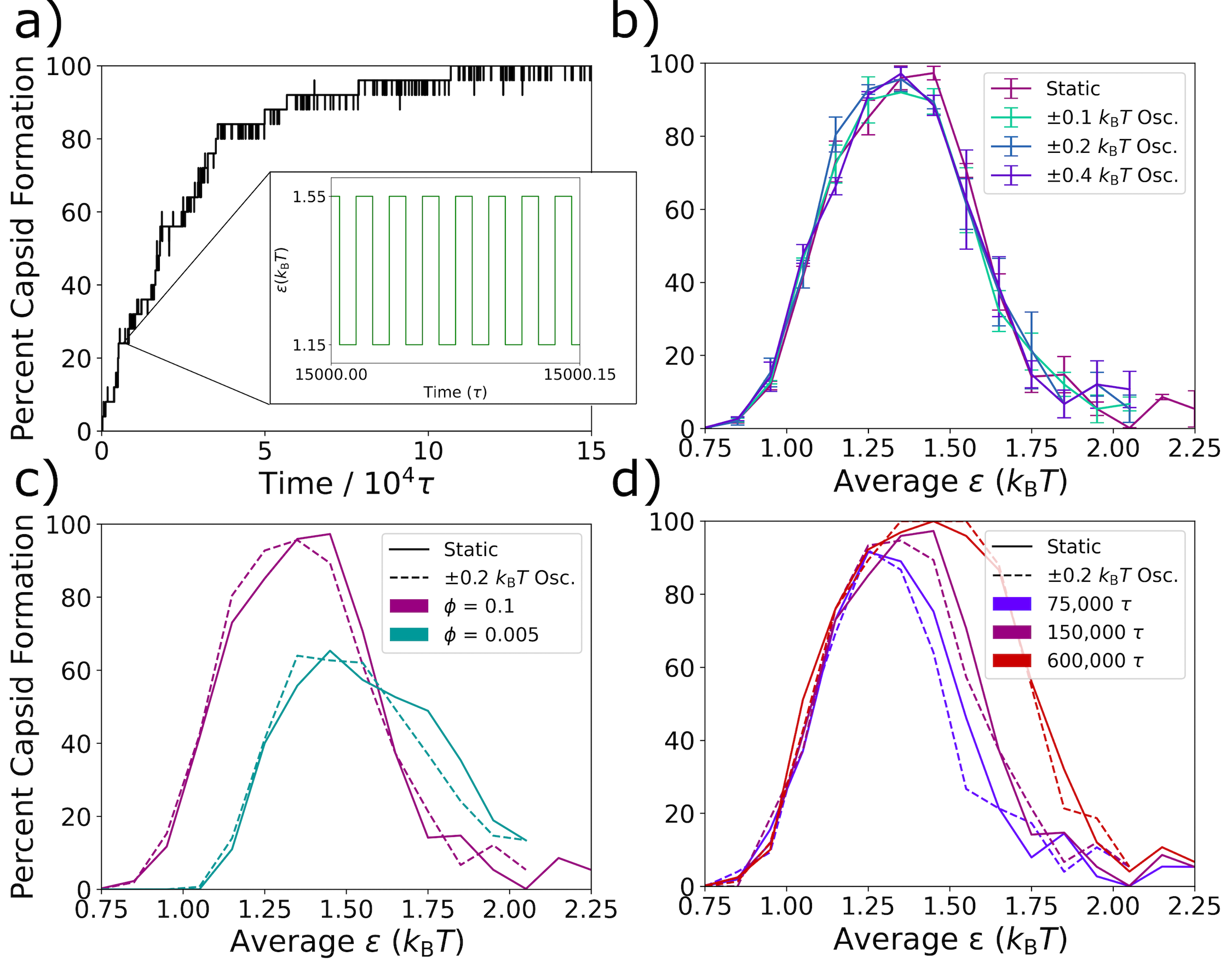}
  \caption{{\bf Assembly at the fast oscillation limit.} (a) A kinetic trace is shown from a single 150,000$\tau$ trajectory of the assembling capsids at an attraction strength of $\epsilon_{\mathrm{avg}} = 1.35k_{\mathrm{B}}T$, an oscillation amplitude of 0.2$k_{\mathrm{B}}T$, and an oscillation period of 0.02$\tau$. Inset displays a zoomed in schematic of the oscillation waveform to compare the oscillation frequency to the timescale of capsid assembly.  
  (b) Capsid yield curves at 150,000$\tau$ are plotted for three different amplitudes as well as the non-oscillatory system at a period of 0.02$\tau$. (c) Capsid yield curves at 150,000$\tau$ are plotted for the oscillatory system at a period of 0.02$\tau$ and the static potential are shown at two different densities. (d) Capsid yield curves at three different simulation times are plotted for the static system (solid lines) and compared to the results for the system with interactions oscillated for a period of 0.02$\tau$ with an amplitude of 0.2$k_{\mathrm{B}}T$ (dashed lines). 
  Percent capsid formation is averaged over three independent trajectories, and the average standard deviations for the capsid yield measurements were $\pm 3.2$\% for the static systems and $\pm 3.8$\% for the oscillatory systems. 
  } 
  
  \label{img:fastlimit}
\end{figure}

To test the the behavior of our system at the fast oscillation limit, in Figure~\ref{img:fastlimit} we compare the degree of capsid formation at static $\epsilon$ values to that at the corresponding $\epsilon_{\mathrm{avg}}$ values for various oscillation amplitudes, monomer concentrations, and simulation times. An oscillation period of 0.02$\tau$ was chosen (see Fig.~\ref{img:fastlimit}a), which is expected to yield fast limit behavior since it is much shorter than the calculated time-scales of interest in Table~\ref{tbl:diffusiontable}. 
In Figure~\ref{img:fastlimit}b, we confirm 
that this period reproduces the static non-monotonic yield curve across three different oscillation amplitudes. 
In Figure~\ref{img:fastlimit}c, we test the correspondence between the static and effective potentials across changes in particle density.  The static yield curves (solid lines) at volume fractions of $\phi = 0.1$ and $\phi = 0.005$ differ, which is expected since a decreased volume fraction requires a stronger $\epsilon$ for capsid formation. However, at both densities, the oscillatory potential with a period of 0.02$\tau$ and an amplitude of 0.2$k_{\mathrm{B}}T$ (dashed lines) returns the corresponding static yield curve for $\epsilon_{\mathrm{avg}}$. 
Finally, in Figure~\ref{img:fastlimit}d, we investigate the timescales of relaxation at the fast oscillation limit as compared to those in the static system by comparing the oscillatory and static capsid yield curves 
after simulation times of 75,000$\tau$, 150,000$\tau$, and 600,000$\tau$. We find that capsid formation at the fast oscillation limit increases with simulation time in precisely the same way as capsid formation does within the corresponding static potential. This result suggests that 
kinetic traps affect the system dynamics at the fast oscillation limit in the same way as in the static system, 
supporting the prior claim that even a system's non-equilibrium dynamics can be described by an appropriate time-averaged potential at the fast oscillation limit\cite{tagliazucchi2016} and explaining the observation that in both Fig.~\ref{img:fastlimit}b and Fig.~\ref{img:fastlimit}c, the correspondence between the oscillatory and static curves is as good for the kinetically constrained right-hand side of the curve as it is for the thermodynamically equilibrated left-hand side. 
In summary, when the Lennard-Jones potential is oscillated at a very short period of 0.02$\tau$, the system organizes in the same manner as with the non-oscillatory potential across a number of tested variations. 

\begin{figure}[h]
        \centering
        \includegraphics[height=6.85cm]{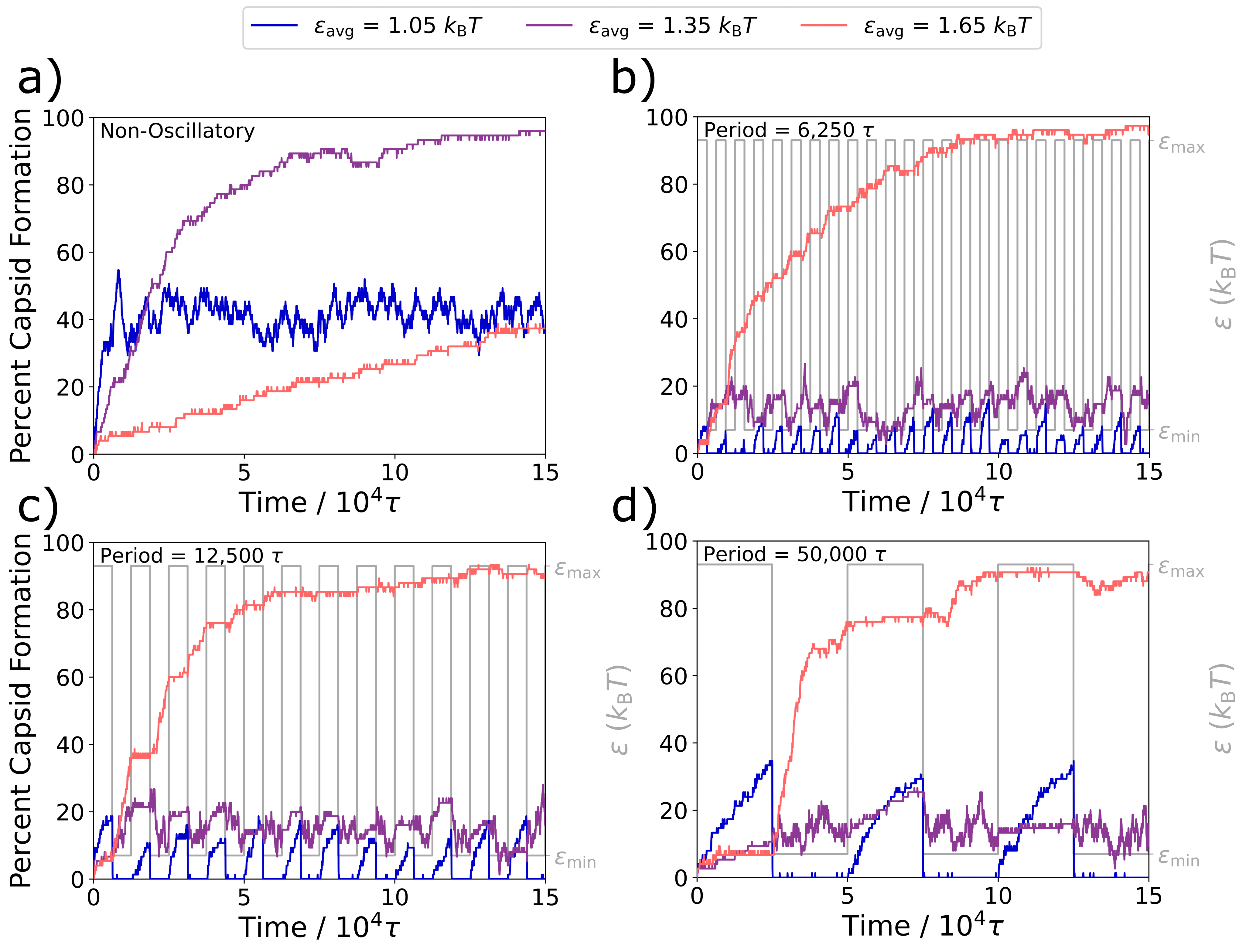}
        \caption{{\bf Capsid formation changes with each half-cycle.} {Capsid formations at different $\epsilon_{\mathrm{avg}}$ values are shown at an oscillation amplitude of 0.4$k_{\mathrm{B}}T$ for three longer periods, as compared to the non-oscillatory capsid formation curves in (a). We show (b) $\tau_{\mathrm{osc}} = 6,250\tau$, (c) $\tau_{\mathrm{osc}} = 12,500\tau$, and (d) $\tau_{\mathrm{osc}} = 50,000\tau$.} Each period has the waveform overlaid in gray to show the oscillation period. 
        The kinetic traces are averaged over three independent trajectories. 
        }
        
        \label{img:Long}
        \end{figure}

{\bf Under slow oscillations, the system adapts to the $\epsilon$ value of each half-cycle.} 
Next we look at the slow oscillation limit, where the period is longer than the displacement times $\{t_{\mathrm{d}}\}$ in Table~\ref{tbl:diffusiontable}, and the system has sufficient time within a single half-cycle to adapt itself substantially to the current $\epsilon$ value.  
Figure~\ref{img:Long} plots capsid yield as a function of simulation time for the static reference case (a) and three longer oscillation periods (b-d), for three different $\epsilon_{\mathrm{avg}}$ values, all with an oscillation amplitude of $0.4k_{\mathrm{B}}T$.

In the Figure~\ref{img:Long}a static case, a weaker attraction strength of $\epsilon_{\mathrm{avg}}= 1.05k_{\mathrm{B}}T$ results in the capsid yield quickly plateauing to about $40\%$, a moderate attraction strength of $\epsilon_{\mathrm{avg}}= 1.35k_{\mathrm{B}}T$ results in nearly complete capsid assembly over a slightly longer time-scale, and a stronger interaction strength of $\epsilon_{\mathrm{avg}}= 1.65k_{\mathrm{B}}T$ results in kinetic trapping and slow capsid assembly.

In Figure~\ref{img:Long}b-d, we plot capsid formation along with the oscillation profiles for periods of 6,250$\tau$, 12,500$\tau$, and 50,000$\tau$, where $\tau_{\mathrm{osc}} >> \{t_{\mathrm{d}}\}$. 
For the weakest interaction strength of $\epsilon_{\mathrm{avg}}= 1.05k_{\mathrm{B}}T$ (in blue), at the given amplitude of 0.4$k_{\mathrm{B}}T$, the simulation  
oscillates between a well-depth of $\epsilon_{\mathrm{min}} = 0.65k_{\mathrm{B}}T$ and $\epsilon_{\mathrm{max}} = 1.45k_{\mathrm{B}}T$. For all three $\tau_{\mathrm{osc}}$ periods, capsids rapidly assemble in the $\epsilon_{\mathrm{max}}$ half-cycle and rapidly disassemble in the $\epsilon_{\mathrm{min}}$ half-cycle, where interactions are too weak for the structures to remain intact.  
As with the weakest $\epsilon_{\mathrm{avg}}$ value, simulations at the strongest $\epsilon_{\mathrm{avg}}$ value investigated here of 1.65 $k_{\mathrm{B}}T$ (in orange) also display significantly different assembly behaviors in each half-cycle. At these stronger attraction strengths, capsids no longer rapidly fall apart during the $\epsilon_{\mathrm{min}}$ half-cycles, however the kinetically trapped snake-like structures can still relax to form additional capsids. As a result, significant capsid formation occurs during the $\epsilon_{\mathrm{min}}$ half-cycles of $\epsilon_{\mathrm{avg}}= 1.65k_{\mathrm{B}}T$. These capsids remain intact during the $\epsilon_{\mathrm{max}}$ half-cycles, however the stronger attractive forces keeping the subparticles together also largely arrest the relaxation of kinetic traps and thus the formation of additional capsids. 
Behavior with the intermediate $\epsilon_{\mathrm{avg}}$ of 1.35 $k_{\mathrm{B}}T$ (in purple) is more complex. The 
system oscillates between $\epsilon_{\mathrm{min}} = 0.95k_{\mathrm{B}}T$ and $\epsilon_{\mathrm{max}} = 1.75k_{\mathrm{B}}T$, both of which result in only modest capsid yields in the static system (see Figure~\ref{img:equilYield}a), and we observe that same modest yield in all three longer oscillation periods in Fig.~\ref{img:Long}b-d, despite the fact that a high yield is obtained for $\epsilon = 1.35 k_{\mathrm{B}}T$ in the static case in Fig.~\ref{img:Long}a.

Overall, when considering oscillations that are slow compared to the characteristic displacement times in  Table~\ref{tbl:diffusiontable}, we see that the system  substantially adapts to each half-cycle, which can result in capsid yields that grow and shrink over each oscillation or in capsid yields that remain static during most $\epsilon_{\mathrm{max}}$ half-cycles and then further increase during most $\epsilon_{\mathrm{min}}$ half-cycle. 
It is important to note that even the very long oscillation periods investigated here are still shorter than the time needed to fully relax the kinetic traps of the system, which can take longer than 75,000$\tau$ even at $\epsilon=1.45~k_{\mathrm{B}}T$, just slightly above the capsid yield curve's peak (see Fig.\ref{img:equilYield}a).

\begin{figure*}
    \centering
    \includegraphics[height=8.8cm]{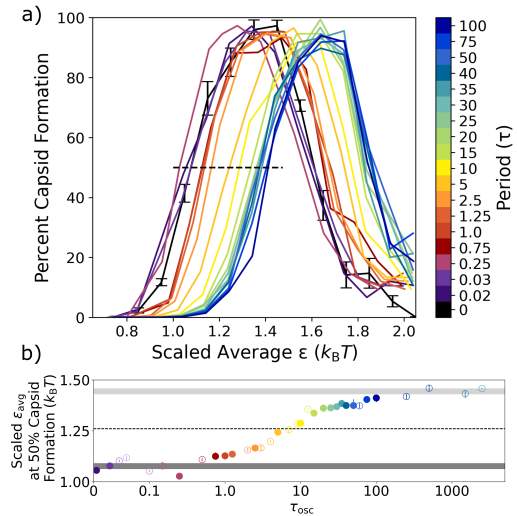}
    \caption{{\bf Capsid yield curves for different periods of oscillation.} 
    (a) Capsid yield curves after 150,000$\tau$ are compared to the scaled $\epsilon_{\mathrm{avg}}$ value (in units of $k_{\rm B}T^{\rm(obs)}$) for different oscillation periods at an amplitude of 0.4$k_{\mathrm{B}}T$. The black, dashed line indicates 50\% capsid formation on the left hand side of the yield curves. (b) The scaled $\epsilon_{\mathrm{avg}}$ values that result in 50\% capsid formation on the left hand side of the capsid yield curve are plotted vs. $\tau_{\mathrm{osc}}$. The dark gray bar indicates the non-oscillatory regime, the light gray bar plots the upper asymptotic limit of a sigmoidal fit, and the dotted line indicates the midpoint between the two.  
    Points in (b) that are shown in (a) are represented by a closed circle, while additional periods not shown in (a) are denoted by an open circle. (c) The percentage of assembled capsids is plotted over 150,000$\tau$ for three different $\epsilon_{\rm avg}$ values ($\epsilon_{\rm avg} = 1.35 k_{\rm B}T$, $\epsilon_{\rm avg} = 1.55 k_{\rm B}T$, $\epsilon_{\rm avg} = 1.75 k_{\rm B}T$) for six different oscillation periods and the static case are shown with the color bar. Capsid yield and kinetic growth curves are averaged over three independent trajectories, and the error bars on the static yield curve show its standard deviation.}
    \label{img:intermediateYield}
    \end{figure*}

{\bf Intermediate oscillation periods causes yield curves to shift to stronger $\epsilon_{avg}$-values.} 
In order to probe how oscillatory interactions influence capsid assembly between the fast and slow oscillation limits, 
in Figure~\ref{img:intermediateYield} we plot the capsid yield curves across a range of oscillation periods at an amplitude of 0.4$k_{\mathrm{B}}T$. 
Details of aggregate formation for three of these periods are shown for a range of $\epsilon_{\mathrm{avg}}$ values in Figure~\ref{fgr:formationTime}. 

For all periods, capsid assembly remains non-monotonic with interaction strength, and the window of orderly assembly is approximately the same width for systems undergoing oscillatory interactions as for those with static interactions. 
However, as the period of oscillation increases in Figure~\ref{img:intermediateYield}a, the capsid yield curves shift to higher $\epsilon_{\mathrm{avg}}$ values, which is similar to results from simulations on colloidal crystal grown under toggled interactions\cite{sherman2018}. Our results show that the window of orderly assembly can be expanded into interaction strengths that are typically infeasible due to their long relaxation times.

One technical complication arises as the oscillation period increases slightly from the fast oscillation limit period of $0.02\tau$. From about 0.03$\tau$ to about $5\tau$, where $\tau_{\rm osc}$ is similar to the timescale of the thermostat relaxation, we find that the Langevin thermostat is not able to maintain the target temperature due to the rate of energy transfer as the interparticle potentials oscillate.
To enable comparisons that include these oscillation frequencies, the $\epsilon_{\mathrm{avg}}$ values in Figure~\ref{img:intermediateYield} were simply scaled  by reporting $\epsilon$ values in units of $k_{\rm B}T^{\rm(obs)}$, where $T^{\rm(obs)}$ is the actual observed temperature at each oscillation period. A plot of the observed temperatures vs. oscillation period is shown in Figure~\ref{fig:temperatures} in the SI.

To quantify the capsid yield curve shift, in Figure~\ref{img:intermediateYield}b we plot the appropriately scaled $\epsilon_{\mathrm{avg}}$ value that results in 50\% capsid formation on the left hand side of the yield curves in Fig.~\ref{img:intermediateYield}a (i.e., where it intersects the black dashed line) vs. the oscillation period, $\tau_{\mathrm{osc}}$.  
The $\epsilon$ value that results in 50\% capsid formation in the static system is indicated by the dark gray bar, while the corresponding  $\epsilon_{\mathrm{avg}}$ value for the long oscillation periods is indicated by the light grey bar.

The result for the fast oscillation limit ($\tau_{\mathrm{osc}}$ = 0.02$\tau$) overlaps with the non-oscillatory dark gray bar, as previously observed in Fig.~\ref{img:fastlimit}. 
For oscillation periods between 0.02$\tau$ and about 0.5$\tau$, the scaled $\epsilon_{\mathrm{avg}}$ values are clustered around the dark gray bar and there is no sustained shift away from the static $\epsilon$ value. After 0.5$\tau$, however, the scaled $\epsilon_{\mathrm{avg}}$ value at 50\% capsid formation steadily increases with oscillation period, crosses the dotted line that indicates the halfway point between the fast and slow oscillation limits at a period of about $\tau_{\mathrm{osc}} =7\tau$, and then plateaus at the slow oscillation value (light grey bar).

To make sense of the shift with oscillation period in Fig.~\ref{img:intermediateYield}, we return to the inherent length-scales of our model and their corresponding local relaxation time-scales, as described in Table~\ref{tbl:diffusiontable}.
After aggregates nucleate, there are three important local relaxation processes that facilitate the second stage of the hexagonal capsid formation from either smaller or larger aggregates: (1) the diffusion of subparticles away from one another; (2) the sliding of triangular particle edges along one another; and (3) the rotation of a triangular particle away from or towards another. All three of these movements are more likely to occur during the $\epsilon_{\mathrm{min}}$ half-cycle when attractions are weaker. 
Table~\ref{tbl:diffusiontable} provides estimates for the timescales of these local relaxation processes, based on simulations of a single triangular particle (see Figure~\ref{img:MSD}).
First, the diffusion of one subparticle out of the LJ attractive well of a neighboring subparticle is characterized by the length-scale of the LJ attraction. The distance a subparticle must move for the attractive interaction to be reduced to a tenth of its full strength is 0.18$\sigma$, which is estimated to take approximately 0.6$\tau$. Second, the edge length of a single triangular particle is 1.0$\sigma$, and the corresponding diffusion time for that distance is 10.3$\tau$. Third, the mean lifetime for the rotational degree of freedom is estimated to be 7.1$\tau$, based on the rotational correlation function calculated in Figure~\ref{img:MSD}. 
These intrinsic local relaxation timescales aid our interpretation of the shift in the capsid yield curves with oscillation period in Fig.~\ref{img:intermediateYield}.

At a period of $\tau_{\mathrm{osc}} =0.5\tau$, where the steady shift towards higher $\epsilon_{\mathrm{avg}}$ values starts, there is just enough time during each $\epsilon_{\mathrm{min}}$ half-period for a subparticle to move away from a neighbor so that their attractive LJ interaction is reduced by two thirds.
At a period of about $1.2\tau$, the $\epsilon_{\mathrm{min}}$ half-cycle is just slightly longer than the time it takes for a subparticle's LJ interaction to be reduced to a tenth of its maximum strength, on average. That is, particles that were originally interacting are likely to diffuse far enough from their original configurations so that they no longer experience a significant attraction to their original neighbors.

The black dashed line indicates the half-way mark between the fast and slow oscillation behaviors, which happens at an oscillation period of about $\tau_{\mathrm{osc}} =7\tau$. From Table~\ref{tbl:diffusiontable}, we see that several local relaxation processes that are helpful in the formation of capsids are achievable within an $\epsilon_{\mathrm{min}}$ half-period of $3.5\tau$. Within this half-cycle, particles can diffuse far enough from their original configurations so that they no longer experience a significant attraction to their original neighbors, even shifting the full diameter of a subparticle. If a particle located in an aggregate can shift a subparticle away from their original configuration, this could lower the energetic barrier for the triangle to diffuse from an aggregate structure.

After a period of 20$\tau$, we observe the start of a plateau in the scaled $\epsilon_{\mathrm{avg}}$ values that corresponds to 50\% capsid formation. Interestingly, a period of 20$\tau$ corresponds to a half-cycle of 10$\tau$, which is about the time needed for a triangular particle to diffuse 1$\sigma$, or the distance of one full edge of the triangle. This edge length is an important distance in disrupting longer aggregates, since a particle in the middle of a snake-like aggregate would need to diffuse about 1$\sigma$ to leave the aggregate and thereby break up a longer aggregate. In addition, this length is central to the motions of a trimer of particles that enable full capsid formation from a half-capsid structure on the end of a snake-like aggregate, as can be seen in Figure~\ref{fgr:structures}. 
Given the mean rotational lifetime of approximately 7$\tau$, it is clear that by an oscillation period of $20\tau$, triangular particles can fully diffuse from aggregate structures and rotate to better align with other particles to facilitate capsid formation during the $\epsilon_{\mathrm{min}}$ half-cycles.

To probe the effect of oscillations on capsid assembly kinetics, in Figure~\ref{img:intermediateYield}c we track the percent capsid formation as a function of simulation time for several oscillation periods and $\epsilon_{\rm avg}$ values. As expected from prior theoretical work,\cite{tagliazucchi2016} the kinetics under the static potential and at the fast oscillation limit are indistinguishable for all $\epsilon_{\rm avg}$ values. In contrast, intermediate oscillation periods result in faster capsid assembly rates relative to the static potential case for all $\epsilon_{\rm avg}$ values. This increase in assembly rates has been previously observed in other oscillatory protocols.\cite{sherman2019,kao2021} We find it is most dramatic at the higher $\epsilon_{\rm avg}$ values (Fig.~\ref{img:intermediateYield}c, bottom two panels) since, at intermediate periods, the oscillations shift the capsid yield curve into the interaction regime where static assembly slows due to kinetic trapping. However, this increased rate holds even in the case of $\epsilon_{\rm avg}$= 1.35$k_{\rm B}T$, where the oscillatory cases with intermediate periods (10$\tau$ and longer) plateau more rapidly than the static case -- even though they plateau at lower capsid yields. 

A closer look at the $\epsilon_{\rm avg}=1.35k_{\rm B}T$ case (Fig.~\ref{img:intermediateYield}c, top panel) provides helpful insight as to why the capsid yield curve shifts to the right as the oscillation period increases but does not broaden. The oscillation amplitude in Fig.~\ref{img:intermediateYield} is 0.4$k_{\rm B}T$, so the $\epsilon_{\rm avg}=1.35k_{\rm B}T$ case oscillates between an $\epsilon_{\rm min}$ value of $0.95k_{\rm B}T$ and an $\epsilon_{\rm max}$ value of $1.75k_{\rm B}T$. At the fast oscillation limit, the capsid yield after 150,000$\tau$ should be equal to that of the static yield at $\epsilon=1.35k_{\rm B}T$ (96\%), which we observe. At the slow oscillation limit, we expect the yield to switch back and forth between the capsid yields expected for the $\epsilon_{\rm min}$ and $\epsilon_{\rm max}$ values, which are 12\% and 14\%, respectively. 
In the intermediate oscillation regime, we see that the system forms a non-equilibrium steady state where the steady-state capsid yield depends on the oscillation period. As the period increases, the capsid yield will eventually transition away from the fast oscillation limit value of 96\% to the slow oscillation limit values of 12\% (an equilibrium value) and 14\% (a kinetically-determined value, which will depend on the initial configuration and the relaxation time). At that slow oscillation limit, it will no longer be in a steady-state, and the number of capsids will shift depending on the oscillation half-cycle. Overall, the observed yield curve on the left-hand side shifts to the right with increased oscillation period. 
On the right-hand side of the shifted curves, when $\epsilon_{\mathrm{min}}$ becomes too strong, the weaker half-cycle can no longer facilitate error correction, which results in the oscillatory capsid yield curves reproducing the static yield curve's non-monotonic behavior.

\subsection{Shift in capsid yield with amplitude provides evidence for the critical role of error correction.} \label{sec:amplitude}

\begin{figure}[H]
\centering
\includegraphics[height=13cm]{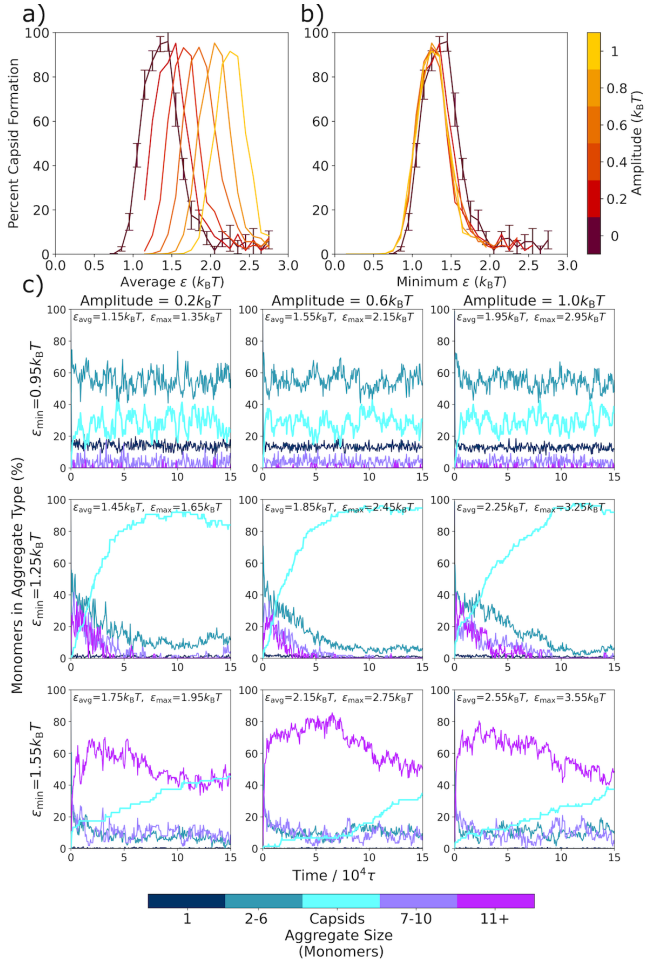}
\caption{{\bf Capsid yield and aggregate formation for different oscillation amplitudes.} 
Capsid yield curves are plotted in (a) vs. $\epsilon_{\mathrm{avg}}$ and in (b) vs. $\epsilon_{\mathrm{min}}$ for five different oscillation amplitudes with an oscillation period of 100$\tau$ and the static potential case. 
The formation of different sized aggregates is plotted vs. time in (c) for three different amplitudes at a period of 100$\tau$. Although they have different $\epsilon_{\mathrm{avg}}$ values, $\epsilon_{\mathrm{min}} = 0.95k_{\mathrm{B}}T$ for all amplitudes in row 1, $\epsilon_{\mathrm{min}} = 1.25k_{\mathrm{B}}T$ for all amplitudes in row 2, and $\epsilon_{\mathrm{min}} = 1.55k_{\mathrm{B}}T$ for all amplitudes in row 3. (d) The capsid growth curves are plotted over 150,000$\tau$ for six different $\epsilon_{\rm min}$ values at five different oscillation amplitudes (shown with the color bar).
Capsid yield curves and monomer counts in the different aggregate types are averaged over three independent trajectories. In the static case in (a) and (b), the error bars represent the standard deviation.}
\label{img:amplitude}
\end{figure}

Previously, Risbud and Swan\cite{risbud2015} and others\cite{promislow1996,promislow1997,swan2012, swan2014, bauer2015, kim2020,martin-roca2023} observed that oscillating attractions could result in local relaxation, since the system was able to relax kinetic barriers via diffusion when the attractions were turned off. 
In our model, the formation of capsid-like structures is also limited by kinetic traps at stronger attractions, as demonstrated in Figure~\ref{img:equilYield}a by the increase in capsid yield with longer simulation times. In addition, at these higher $\epsilon$ values, the initial formation of snake-like structures that can convert into capsid-like hexamers makes it clear that error correction -- the ability of sub-optimally assembled particles to rearrange themselves into a more favorable structure -- plays an important role in the total capsid yield, both in the static and oscillatory interaction cases.
Indeed, in Figure~\ref{img:Long}b-d, at the slow oscillation limit with an amplitude of 0.4$k_{\mathrm{B}}T$, we directly observed the effect of error correction on the capsid formation process during the lower $\epsilon_{\mathrm{min}}$ half-cycles for the case where $\epsilon_{\mathrm{avg}}= 1.65k_{\mathrm{B}}T$. Capsid yields almost always remained static in the $\epsilon_{\mathrm{max}}$ half-cycles, but ratcheted up to higher levels in the $\epsilon_{\mathrm{min}}$ half-cycles. 
In this section, we further probe how the $\epsilon_{\mathrm{min}}$ half-cycles affect error correction within this model system by varying the oscillation amplitude.
As the amplitude increases, $\epsilon_{\mathrm{avg}}$ stays the same, but attractions oscillate between a weaker $\epsilon_{\mathrm{min}}$ and stronger $\epsilon_{\mathrm{max}}$. Results are shown in Figure~\ref{img:amplitude} for an intermediate oscillation period of 100$\tau$, which is long enough for most local relaxation processes to occur -- see Table~\ref{tbl:diffusiontable} and the almost-completed shift towards the long-time oscillation value at 100$\tau$ in Figure~\ref{img:intermediateYield}b.

In Figure~\ref{img:amplitude}a, we show a series of capsid yield curves at different oscillation amplitudes plotted vs. $\epsilon_{\mathrm{avg}}$.  
As amplitude increases, yield curves shift to the right to higher $\epsilon_{\mathrm{avg}}$ values -- values where orderly assembly is not observed in the static system. 
However, when we plot these same yield curves vs. $\epsilon_{\mathrm{min}}$ instead of $\epsilon_{\mathrm{avg}}$ in Figure~\ref{img:amplitude}b, the curves for all amplitudes collapse into one curve, which is very close to the capsid yield curve for the static potential.
This collapse of the yield curves indicates that, at oscillation periods sufficiently longer than the local intrinsic relaxation processes in the system, the main determinant of capsid yield in the oscillatory system is simply the value of $\epsilon_{\mathrm{min}}$. 
Indeed, this full shift from the yields observed at $\epsilon_{\mathrm{avg}}$ to those observed at $\epsilon_{\mathrm{min}}$ as the oscillation periods lengthen can be clearly seen in Figure~\ref{img:intermediateYield}a.

To further probe this shift to the $\epsilon_{\mathrm{min}}$ value fully determining the yield at oscillation times longer than the local relaxation processes, in Figure~\ref{img:amplitude}c we show the percent of the triangular monomers in each aggregate size as a function of time for a series of simulations with varying amplitudes and $\epsilon_{\mathrm{min}}$ values. From the top row to the bottom row, $\epsilon_{\mathrm{min}}$ increases, while within each row, $\epsilon_{\mathrm{min}}$ is held fixed while the amplitude increases from left to right.
Notably, we find that the kinetic assembly traces for a given $\epsilon_{\mathrm{min}}$ value are essentially the same across the three amplitudes. In Figure~\ref{img:amplitude}d we overlay the kinetic traces for full capsid assembly over more amplitudes and $\epsilon_{\mathrm{min}}$ values. Regardless of oscillation amplitude, the capsid yield curves collapse onto one another for each $\epsilon_{\rm min}$ value, demonstrating that $\epsilon_{\rm min}$ largely determines capsid assembly even as $\epsilon_{\rm avg}$ changes. Similar kinetic curve collapses have been observed in experiments of polarizable colloids exposed to toggled magnetic fields\cite{swan2012} and simulations of nanoparticles with toggled potentials\cite{sherman2018}.
This universal $\epsilon_{\mathrm{min}}$ dependence shows that, as long as oscillation periods are longer than the local intrinsic relaxation processes and $\epsilon_{\mathrm{max}}$ is above the threshold for kinetic trap formation, the strength of $\epsilon_{\mathrm{min}}$ is what determines the degree to which kinetically trapped aggregates are able to relax into fully formed capsids.

\begin{figure}[h]
\centering
\includegraphics[height=5.25cm]{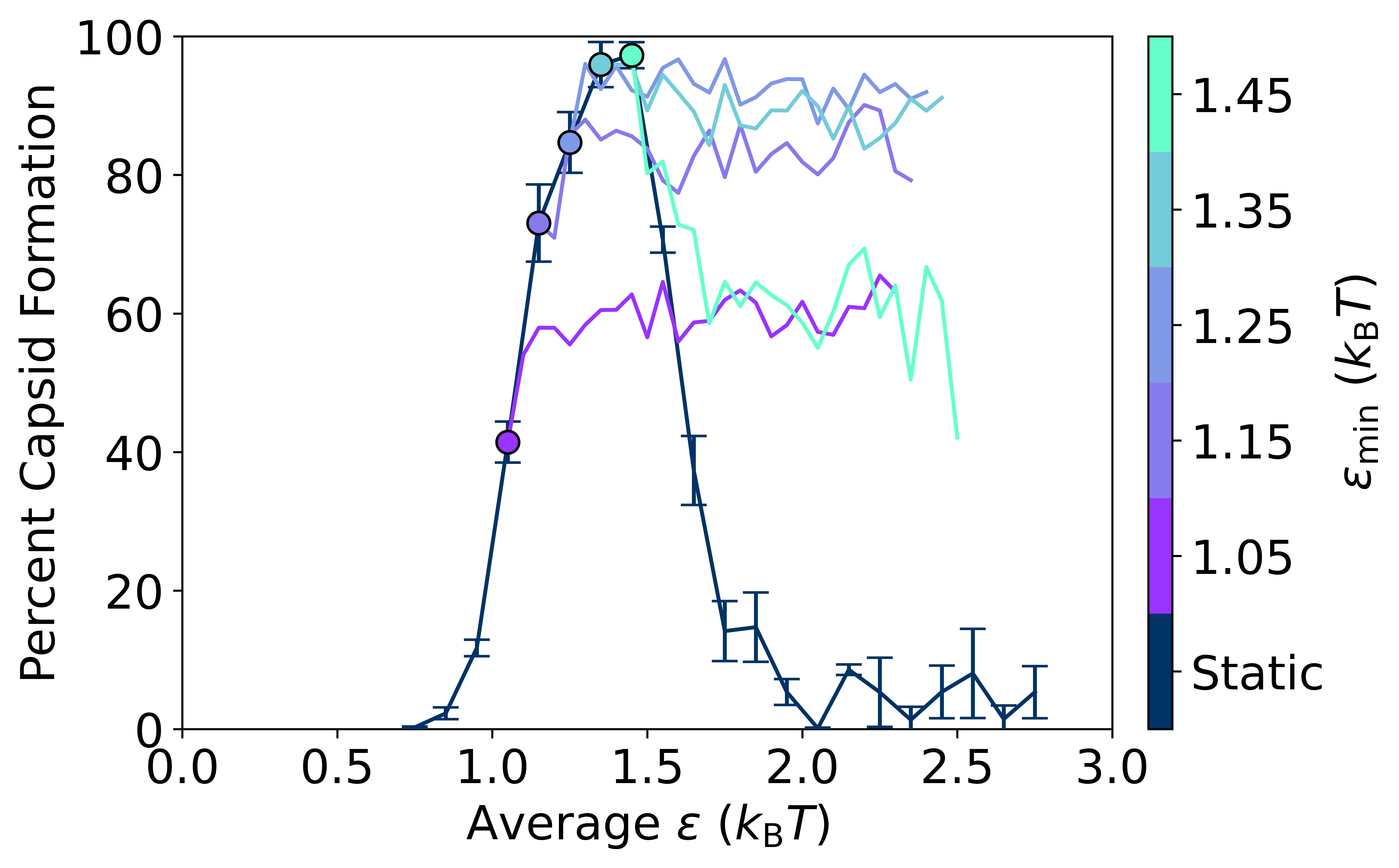}
\caption{{\bf Capsid yield for oscillation protocols where $\epsilon_{\rm min}$ is held constant.} Capsid yields are plotted at an oscillation period of 100$\tau$ for five fixed $\epsilon_{\rm min}$ conditions while $\epsilon_{\rm max}$ is varied, and the results are plotted vs. $\epsilon_{\rm avg}$. The static yield curve is shown for comparison, and the static result corresponding to each $\epsilon_{\rm min}$ series is indicated with the solid colored dot. All capsid yields plotted have been averaged over three independent trajectories. The error bars show the standard deviations on the static curve.}
\label{img:broaden}
\end{figure}

Building on this insight, in Figure~\ref{img:broaden}, we plot capsid yield curves for static and oscillatory protocols where $\epsilon_{\rm min}$ is fixed at different attraction strengths, $\epsilon_{\rm max}$ is varied, and the results are plotted vs. $\epsilon_{\rm avg}$.
On the left-hand size, there is a slight increase in the yield as the $\epsilon_{\rm avg}$ value increases from the corresponding static case (indicated by the dot). In contrast, on the right-hand side, there is a drop off in the yield as $\epsilon_{\rm avg}$ increases from the static case. 
In both cases, the yield appears to plateau at some point as $\epsilon_{\rm avg}$ increases further. Essentially, further increases in attraction strength past some value of $\epsilon_{\rm max}$ do not change the resulting yield, which then depends only on the error correction enabled by the $\epsilon_{\rm min}$ value, as shown in Figure~\ref{img:amplitude}. As a result, fixing $\epsilon_{\rm min}$ to a value just below the peak of the static capsid yield curve removes the non-monotonicity of the static yield curve.

\section{Conclusion} \label{sec:conc}

In this work, we investigate how temporal oscillations influence the dissipative self-assembly of anisotropic 2D triangular particles that assemble into self-limited capsid-like structures. At stronger attractions, the formation of these hexamers occurs through a non-classical, two-step nucleation pathway and is prone to kinetic trapping, resulting in a non-monotonic dependence of capsid yield on attraction strength in both the static and oscillatory cases. 

Temporal oscillations in inter-particle attractions offer experimentally accessible dissipative self-assembly protocols for a wide range of materials. Such oscillations may be implemented by changes in temperature, external applied fields, light irradiation, pH changes, or other mechanisms, and in our study we implement them by changing the strength of the attractions in time by switching between two LJ well-depths. 
As in previous studies, we find that tuning the oscillations enables orderly assembly across a broader range of time-averaged interactions, as can be seen in the envelope of capsid yield curves in Figure~\ref{img:intermediateYield}
and the full region of $\epsilon_{\rm avg}$ that they span, which provides routes to substantive capsid yields at any of these time-averaged interaction strengths. In addition, we generally observe faster capsid formation under oscillatory conditions, providing a route to more rapidly produce self-limited assembled structures.\cite{swan2014,sherman2018,kao2021}

In keeping with prior work,\cite{risbud2015,kao2021} our results highlight the correspondence between the different oscillation timescales and the particle motions that govern local relaxation processes -- in the case of our more complex anisotropic particles, these include both translations and rotations. At the limit of oscillations that are very fast compared to these intrinsic time-scales, assembly proceeds as if the system were subject to the static attraction that is the time-averaged strength over a single oscillation period, defined by the averaged LJ well-depth, $\epsilon_{\rm avg}$. At the slow oscillation limit, the system evolves according to the attractive forces at play within the current half-cycle. In between these extremes, the assembly yield curve shifts from one determined by $\epsilon_{\rm avg}$ to one determined by $\epsilon_{\rm min}$, the attraction strength during the weaker attraction half-cycles, since it is the error correction made possible at those times that determines the degree to which kinetic traps can be overcome.

These findings provide insights that could aid in the design of time-dependent assembly protocols for novel materials. 
Advanced materials may be composed of more complex, hierarchically ordered components with multiple intrinsic lengths, and thus multiple relaxation time-scales. Even without closed feedback loops or detailed simulations of the specific system of interest, we show that designing protocols with interactions that oscillate over times similar to the intrinsic relaxation times of that material may provide efficient assembly routes.  Estimating the translational and rotational timescales that corresponds to important energetic and structural length scales within the assembling system will serve as a starting point for creating oscillation protocols that can relax kinetically trapped structures. In addition, the choice of an oscillation amplitude may be guided by evaluating the interaction regimes where moderate assembly occurs under static conditions, as we find that error correction within a system can be enhanced by periodically switching to attraction strengths that promote a moderate yield and are therefore weak enough to allow critical relaxation processes to proceed.

\section*{Author Contributions}
Jessica K. Niblo: conceptualization, investigation, methodology, data curation, software, formal analysis, visualization, writing – original draft. 
Jacob R. Swartley: methodology, writing – reviewing. 
Zhongmin Zhang: methodology, writing – reviewing. 
Kateri H. DuBay: conceptualization, methodology, supervision, formal analysis, writing – original draft.

\section*{Conflicts of interest}
There are no conflicts to declare.

\section*{Acknowledgements}
This material is based upon work supported by a startup grant from the University of Virginia. The authors also acknowledge Research Computing at the University of Virginia (https://rc.virginia.edu) for providing computational resources and technical support.

\clearpage

\normalem
\bibliography{main}

\providecommand{\latin}[1]{#1}
\makeatletter
\providecommand{\doi}
  {\begingroup\let\do\@makeother\dospecials
  \catcode`\{=1 \catcode`\}=2 \doi@aux}
\providecommand{\doi@aux}[1]{\endgroup\texttt{#1}}
\makeatother
\providecommand*\mcitethebibliography{\thebibliography}
\csname @ifundefined\endcsname{endmcitethebibliography}  {\let\endmcitethebibliography\endthebibliography}{}
\begin{mcitethebibliography}{68}
\providecommand*\natexlab[1]{#1}
\providecommand*\mciteSetBstSublistMode[1]{}
\providecommand*\mciteSetBstMaxWidthForm[2]{}
\providecommand*\mciteBstWouldAddEndPuncttrue
  {\def\EndOfBibitem{\unskip.}}
\providecommand*\mciteBstWouldAddEndPunctfalse
  {\let\EndOfBibitem\relax}
\providecommand*\mciteSetBstMidEndSepPunct[3]{}
\providecommand*\mciteSetBstSublistLabelBeginEnd[3]{}
\providecommand*\EndOfBibitem{}
\mciteSetBstSublistMode{f}
\mciteSetBstMaxWidthForm{subitem}{(\alph{mcitesubitemcount})}
\mciteSetBstSublistLabelBeginEnd
  {\mcitemaxwidthsubitemform\space}
  {\relax}
  {\relax}

\bibitem[Cho \latin{et~al.}(2013)Cho, Hwang, Solandt, and Frank]{cho2013a}
Cho,~N.-J.; Hwang,~L.~Y.; Solandt,~J. J.~R.; Frank,~C.~W. Comparison of {{Extruded}} and {{Sonicated Vesicles}} for {{Planar Bilayer Self-Assembly}}. \emph{Materials} \textbf{2013}, \emph{6}, 3294--3308\relax
\mciteBstWouldAddEndPuncttrue
\mciteSetBstMidEndSepPunct{\mcitedefaultmidpunct}
{\mcitedefaultendpunct}{\mcitedefaultseppunct}\relax
\EndOfBibitem
\bibitem[Zlotnick(1994)]{zlotnick1994}
Zlotnick,~A. To {{Build}} a {{Virus Capsid}} : {{An Equilibrium Model}} of the {{Self Assembly}} of {{Polyhedral Protein Complexes}}. \emph{The Journal of Molecular Biology} \textbf{1994}, \emph{241}, 59--67\relax
\mciteBstWouldAddEndPuncttrue
\mciteSetBstMidEndSepPunct{\mcitedefaultmidpunct}
{\mcitedefaultendpunct}{\mcitedefaultseppunct}\relax
\EndOfBibitem
\bibitem[Zlotnick \latin{et~al.}(2000)Zlotnick, Aldrich, Johnson, Ceres, and Young]{zlotnick2000}
Zlotnick,~A.; Aldrich,~R.; Johnson,~J.~M.; Ceres,~P.; Young,~M.~J. Mechanism of {{Capsid Assembly}} for an {{Icosahedral Plant Virus}}. \emph{Virology} \textbf{2000}, \emph{277}, 450--456\relax
\mciteBstWouldAddEndPuncttrue
\mciteSetBstMidEndSepPunct{\mcitedefaultmidpunct}
{\mcitedefaultendpunct}{\mcitedefaultseppunct}\relax
\EndOfBibitem
\bibitem[Hicks and Henley(2006)Hicks, and Henley]{hicks2006}
Hicks,~S.~D.; Henley,~C.~L. Irreversible Growth Model for Virus Capsid Assembly. \emph{Physical Review E} \textbf{2006}, \emph{74}, 031912\relax
\mciteBstWouldAddEndPuncttrue
\mciteSetBstMidEndSepPunct{\mcitedefaultmidpunct}
{\mcitedefaultendpunct}{\mcitedefaultseppunct}\relax
\EndOfBibitem
\bibitem[Hagan and Chandler(2006)Hagan, and Chandler]{hagan2006}
Hagan,~M.~F.; Chandler,~D. Dynamic {{Pathways}} for {{Viral Capsid Assembly}}. \emph{Biophysical Journal} \textbf{2006}, \emph{91}, 42--54\relax
\mciteBstWouldAddEndPuncttrue
\mciteSetBstMidEndSepPunct{\mcitedefaultmidpunct}
{\mcitedefaultendpunct}{\mcitedefaultseppunct}\relax
\EndOfBibitem
\bibitem[Sigl \latin{et~al.}(2021)Sigl, Willner, Engelen, Kretzmann, Sachenbacher, Liedl, Kolbe, Wilsch, Aghvami, Protzer, Hagan, Fraden, and Dietz]{sigl2021}
Sigl,~C.; Willner,~E.~M.; Engelen,~W.; Kretzmann,~J.~A.; Sachenbacher,~K.; Liedl,~A.; Kolbe,~F.; Wilsch,~F.; Aghvami,~S.~A.; Protzer,~U.; Hagan,~M.~F.; Fraden,~S.; Dietz,~H. Programmable Icosahedral Shell System for Virus Trapping. \emph{Nature Materials} \textbf{2021}, \emph{20}, 1281--1289\relax
\mciteBstWouldAddEndPuncttrue
\mciteSetBstMidEndSepPunct{\mcitedefaultmidpunct}
{\mcitedefaultendpunct}{\mcitedefaultseppunct}\relax
\EndOfBibitem
\bibitem[Lynd \latin{et~al.}(2008)Lynd, Meuler, and Hillmyer]{lynd2008}
Lynd,~N.~A.; Meuler,~A.~J.; Hillmyer,~M.~A. Polydispersity and Block Copolymer Self-Assembly. \emph{Progress in Polymer Science} \textbf{2008}, \emph{33}, 875--893\relax
\mciteBstWouldAddEndPuncttrue
\mciteSetBstMidEndSepPunct{\mcitedefaultmidpunct}
{\mcitedefaultendpunct}{\mcitedefaultseppunct}\relax
\EndOfBibitem
\bibitem[Zhang and Glotzer(2004)Zhang, and Glotzer]{zhang2004}
Zhang,~Z.; Glotzer,~S.~C. Self-{{Assembly}} of {{Patchy Particles}}. \emph{Nano Letters} \textbf{2004}, \emph{4}, 1407--1413\relax
\mciteBstWouldAddEndPuncttrue
\mciteSetBstMidEndSepPunct{\mcitedefaultmidpunct}
{\mcitedefaultendpunct}{\mcitedefaultseppunct}\relax
\EndOfBibitem
\bibitem[Glotzer and Solomon(2007)Glotzer, and Solomon]{glotzer2007}
Glotzer,~S.~C.; Solomon,~M.~J. Anisotropy of Building Blocks and Their Assembly into Complex Structures. \emph{Nature Materials} \textbf{2007}, \emph{6}, 557--562\relax
\mciteBstWouldAddEndPuncttrue
\mciteSetBstMidEndSepPunct{\mcitedefaultmidpunct}
{\mcitedefaultendpunct}{\mcitedefaultseppunct}\relax
\EndOfBibitem
\bibitem[Hormoz and Brenner(2011)Hormoz, and Brenner]{hormoz2011}
Hormoz,~S.; Brenner,~M.~P. Design Principles for Self-Assembly with Short-Range Interactions. \emph{Proceedings of the National Academy of Sciences} \textbf{2011}, \emph{108}, 5193--5198\relax
\mciteBstWouldAddEndPuncttrue
\mciteSetBstMidEndSepPunct{\mcitedefaultmidpunct}
{\mcitedefaultendpunct}{\mcitedefaultseppunct}\relax
\EndOfBibitem
\bibitem[Haxton and Whitelam(2012)Haxton, and Whitelam]{haxton2012}
Haxton,~T.~K.; Whitelam,~S. Design Rules for the Self-Assembly of a Protein Crystal. \emph{Soft Matter} \textbf{2012}, \emph{8}, 3558\relax
\mciteBstWouldAddEndPuncttrue
\mciteSetBstMidEndSepPunct{\mcitedefaultmidpunct}
{\mcitedefaultendpunct}{\mcitedefaultseppunct}\relax
\EndOfBibitem
\bibitem[Gr{\"u}nwald and Geissler(2014)Gr{\"u}nwald, and Geissler]{grunwald2014}
Gr{\"u}nwald,~M.; Geissler,~P.~L. Patterns without {{Patches}}: {{Hierarchical Self-Assembly}} of {{Complex Structures}} from {{Simple Building Blocks}}. \emph{ACS Nano} \textbf{2014}, \emph{8}, 5891--5897\relax
\mciteBstWouldAddEndPuncttrue
\mciteSetBstMidEndSepPunct{\mcitedefaultmidpunct}
{\mcitedefaultendpunct}{\mcitedefaultseppunct}\relax
\EndOfBibitem
\bibitem[Mallory and Cacciuto(2016)Mallory, and Cacciuto]{mallory2016}
Mallory,~S.~A.; Cacciuto,~A. Activity-Assisted Self-Assembly of Colloidal Particles. \emph{Physical Review E} \textbf{2016}, \emph{94}, 022607\relax
\mciteBstWouldAddEndPuncttrue
\mciteSetBstMidEndSepPunct{\mcitedefaultmidpunct}
{\mcitedefaultendpunct}{\mcitedefaultseppunct}\relax
\EndOfBibitem
\bibitem[Damasceno \latin{et~al.}(2012)Damasceno, Engel, and Glotzer]{damasceno2012}
Damasceno,~P.~F.; Engel,~M.; Glotzer,~S.~C. Predictive {{Self-Assembly}} of {{Polyhedra}} into {{Complex Structures}}. \emph{Science} \textbf{2012}, \emph{337}, 453--457\relax
\mciteBstWouldAddEndPuncttrue
\mciteSetBstMidEndSepPunct{\mcitedefaultmidpunct}
{\mcitedefaultendpunct}{\mcitedefaultseppunct}\relax
\EndOfBibitem
\bibitem[Hagan and Elrad(2010)Hagan, and Elrad]{hagan2010}
Hagan,~M.~F.; Elrad,~O.~M. Understanding the {{Concentration Dependence}} of {{Viral Capsid Assembly Kinetics}}{\textemdash}the {{Origin}} of the {{Lag Time}} and {{Identifying}} the {{Critical Nucleus Size}}. \emph{Biophysical Journal} \textbf{2010}, \emph{98}, 1065--1074\relax
\mciteBstWouldAddEndPuncttrue
\mciteSetBstMidEndSepPunct{\mcitedefaultmidpunct}
{\mcitedefaultendpunct}{\mcitedefaultseppunct}\relax
\EndOfBibitem
\bibitem[Grant \latin{et~al.}(2011)Grant, Jack, and Whitelam]{grant2011}
Grant,~J.; Jack,~R.~L.; Whitelam,~S. Analyzing Mechanisms and Microscopic Reversibility of Self-Assembly. \emph{The Journal of Chemical Physics} \textbf{2011}, \emph{135}, 214505\relax
\mciteBstWouldAddEndPuncttrue
\mciteSetBstMidEndSepPunct{\mcitedefaultmidpunct}
{\mcitedefaultendpunct}{\mcitedefaultseppunct}\relax
\EndOfBibitem
\bibitem[Pawar and Kretzschmar(2010)Pawar, and Kretzschmar]{pawar2010}
Pawar,~A.~B.; Kretzschmar,~I. Fabrication, {{Assembly}}, and {{Application}} of {{Patchy Particles}}. \emph{Macromolecular Rapid Communications} \textbf{2010}, \emph{31}, 150--168\relax
\mciteBstWouldAddEndPuncttrue
\mciteSetBstMidEndSepPunct{\mcitedefaultmidpunct}
{\mcitedefaultendpunct}{\mcitedefaultseppunct}\relax
\EndOfBibitem
\bibitem[Whitelam and Jack(2015)Whitelam, and Jack]{whitelam2015a}
Whitelam,~S.; Jack,~R.~L. The {{Statistical Mechanics}} of {{Dynamic Pathways}} to {{Self-Assembly}}. \emph{Annual Review of Physical Chemistry} \textbf{2015}, \emph{66}, 143--163\relax
\mciteBstWouldAddEndPuncttrue
\mciteSetBstMidEndSepPunct{\mcitedefaultmidpunct}
{\mcitedefaultendpunct}{\mcitedefaultseppunct}\relax
\EndOfBibitem
\bibitem[Hagan \latin{et~al.}(2011)Hagan, Elrad, and Jack]{hagan2011}
Hagan,~M.~F.; Elrad,~O.~M.; Jack,~R.~L. Mechanisms of Kinetic Trapping in Self-Assembly and Phase Transformation. \emph{The Journal of Chemical Physics} \textbf{2011}, \emph{135}, 104115\relax
\mciteBstWouldAddEndPuncttrue
\mciteSetBstMidEndSepPunct{\mcitedefaultmidpunct}
{\mcitedefaultendpunct}{\mcitedefaultseppunct}\relax
\EndOfBibitem
\bibitem[Goodrich \latin{et~al.}(2021)Goodrich, King, Schoenholz, Cubuk, and Brenner]{goodrich2021}
Goodrich,~C.~P.; King,~E.~M.; Schoenholz,~S.~S.; Cubuk,~E.~D.; Brenner,~M.~P. Designing Self-Assembling Kinetics with Differentiable Statistical Physics Models. \emph{Proceedings of the National Academy of Sciences of the United States of America} \textbf{2021}, \emph{118}, e2024083118\relax
\mciteBstWouldAddEndPuncttrue
\mciteSetBstMidEndSepPunct{\mcitedefaultmidpunct}
{\mcitedefaultendpunct}{\mcitedefaultseppunct}\relax
\EndOfBibitem
\bibitem[Whitesides and Boncheva(2002)Whitesides, and Boncheva]{whitesides2002a}
Whitesides,~G.~M.; Boncheva,~M. Beyond Molecules: {{Self-assembly}} of Mesoscopic and Macroscopic Components. \emph{Proceedings of the National Academy of Sciences} \textbf{2002}, \emph{99}, 4769--4774\relax
\mciteBstWouldAddEndPuncttrue
\mciteSetBstMidEndSepPunct{\mcitedefaultmidpunct}
{\mcitedefaultendpunct}{\mcitedefaultseppunct}\relax
\EndOfBibitem
\bibitem[Fialkowski \latin{et~al.}(2006)Fialkowski, Bishop, Klajn, Smoukov, Campbell, and Grzybowski]{fialkowski2006}
Fialkowski,~M.; Bishop,~K. J.~M.; Klajn,~R.; Smoukov,~S.~K.; Campbell,~C.~J.; Grzybowski,~B.~A. Principles and {{Implementations}} of {{Dissipative}} ({{Dynamic}}) {{Self-Assembly}}. \emph{The Journal of Physical Chemistry B} \textbf{2006}, \emph{110}, 2482--2496\relax
\mciteBstWouldAddEndPuncttrue
\mciteSetBstMidEndSepPunct{\mcitedefaultmidpunct}
{\mcitedefaultendpunct}{\mcitedefaultseppunct}\relax
\EndOfBibitem
\bibitem[Ilse \latin{et~al.}(2016)Ilse, Holm, and {de Graaf}]{ilse2016}
Ilse,~S.~E.; Holm,~C.; {de Graaf},~J. Surface Roughness Stabilizes the Clustering of Self-Propelled Triangles. \emph{The Journal of Chemical Physics} \textbf{2016}, \emph{145}, 134904\relax
\mciteBstWouldAddEndPuncttrue
\mciteSetBstMidEndSepPunct{\mcitedefaultmidpunct}
{\mcitedefaultendpunct}{\mcitedefaultseppunct}\relax
\EndOfBibitem
\bibitem[L{\"o}wen(2018)]{lowen2018}
L{\"o}wen,~H. Active Colloidal Molecules. \emph{Europhysics Letters} \textbf{2018}, \emph{121}, 58001\relax
\mciteBstWouldAddEndPuncttrue
\mciteSetBstMidEndSepPunct{\mcitedefaultmidpunct}
{\mcitedefaultendpunct}{\mcitedefaultseppunct}\relax
\EndOfBibitem
\bibitem[Das and Limmer(2021)Das, and Limmer]{das2021b}
Das,~A.; Limmer,~D.~T. Variational Design Principles for Nonequilibrium Colloidal Assembly. \emph{The Journal of Chemical Physics} \textbf{2021}, \emph{154}, 014107\relax
\mciteBstWouldAddEndPuncttrue
\mciteSetBstMidEndSepPunct{\mcitedefaultmidpunct}
{\mcitedefaultendpunct}{\mcitedefaultseppunct}\relax
\EndOfBibitem
\bibitem[Das and Limmer(2023)Das, and Limmer]{das2023}
Das,~A.; Limmer,~D.~T. Nonequilibrium Design Strategies for Functional Colloidal Assemblies. \emph{Proceedings of the National Academy of Sciences} \textbf{2023}, \emph{120}, e2217242120\relax
\mciteBstWouldAddEndPuncttrue
\mciteSetBstMidEndSepPunct{\mcitedefaultmidpunct}
{\mcitedefaultendpunct}{\mcitedefaultseppunct}\relax
\EndOfBibitem
\bibitem[Tagliazucchi \latin{et~al.}(2014)Tagliazucchi, Weiss, and Szleifer]{tagliazucchi2014}
Tagliazucchi,~M.; Weiss,~E.~A.; Szleifer,~I. Dissipative Self-Assembly of Particles Interacting through Time-Oscillatory Potentials. \emph{Proceedings of the National Academy of Sciences} \textbf{2014}, \emph{111}, 9751--9756\relax
\mciteBstWouldAddEndPuncttrue
\mciteSetBstMidEndSepPunct{\mcitedefaultmidpunct}
{\mcitedefaultendpunct}{\mcitedefaultseppunct}\relax
\EndOfBibitem
\bibitem[Tagliazucchi and Szleifer(2016)Tagliazucchi, and Szleifer]{tagliazucchi2016}
Tagliazucchi,~M.; Szleifer,~I. Dynamics of Dissipative Self-Assembly of Particles Interacting through Oscillatory Forces. \emph{Faraday Discussions} \textbf{2016}, \emph{186}, 399--418\relax
\mciteBstWouldAddEndPuncttrue
\mciteSetBstMidEndSepPunct{\mcitedefaultmidpunct}
{\mcitedefaultendpunct}{\mcitedefaultseppunct}\relax
\EndOfBibitem
\bibitem[Long \latin{et~al.}(2018)Long, Lei, Ren, and Ma]{long2018}
Long,~C.; Lei,~Q.-l.; Ren,~C.-l.; Ma,~Y.-q. Three-{{Dimensional Non-Close-Packed Structures}} of {{Oppositely Charged Colloids Driven}} by {{pH Oscillation}}. \emph{The Journal of Physical Chemistry B} \textbf{2018}, \emph{122}, 3196--3201\relax
\mciteBstWouldAddEndPuncttrue
\mciteSetBstMidEndSepPunct{\mcitedefaultmidpunct}
{\mcitedefaultendpunct}{\mcitedefaultseppunct}\relax
\EndOfBibitem
\bibitem[Promislow and Gast(1996)Promislow, and Gast]{promislow1996}
Promislow,~J. H.~E.; Gast,~A.~P. Magnetorheological {{Fluid Structure}} in a {{Pulsed Magnetic Field}}. \emph{Langmuir} \textbf{1996}, \emph{12}, 4095--4102\relax
\mciteBstWouldAddEndPuncttrue
\mciteSetBstMidEndSepPunct{\mcitedefaultmidpunct}
{\mcitedefaultendpunct}{\mcitedefaultseppunct}\relax
\EndOfBibitem
\bibitem[Promislow and Gast(1997)Promislow, and Gast]{promislow1997}
Promislow,~J. H.~E.; Gast,~A.~P. Low-Energy Suspension Structure of a Magnetorheological Fluid. \emph{Physical Review E} \textbf{1997}, \emph{56}, 642--651\relax
\mciteBstWouldAddEndPuncttrue
\mciteSetBstMidEndSepPunct{\mcitedefaultmidpunct}
{\mcitedefaultendpunct}{\mcitedefaultseppunct}\relax
\EndOfBibitem
\bibitem[Swan \latin{et~al.}(2012)Swan, Vasquez, Whitson, Fincke, Wakata, Magnus, De~Winne, Barratt, Agui, Green, Hall, Bohman, Bunnell, Gast, and Furst]{swan2012}
Swan,~J.~W.; Vasquez,~P.~A.; Whitson,~P.~A.; Fincke,~E.~M.; Wakata,~K.; Magnus,~S.~H.; De~Winne,~F.; Barratt,~M.~R.; Agui,~J.~H.; Green,~R.~D.; Hall,~N.~R.; Bohman,~D.~Y.; Bunnell,~C.~T.; Gast,~A.~P.; Furst,~E.~M. Multi-Scale Kinetics of a Field-Directed Colloidal Phase Transition. \emph{Proceedings of the National Academy of Sciences} \textbf{2012}, \emph{109}, 16023--16028\relax
\mciteBstWouldAddEndPuncttrue
\mciteSetBstMidEndSepPunct{\mcitedefaultmidpunct}
{\mcitedefaultendpunct}{\mcitedefaultseppunct}\relax
\EndOfBibitem
\bibitem[Swan \latin{et~al.}(2014)Swan, Bauer, Liu, and Furst]{swan2014}
Swan,~J.~W.; Bauer,~J.~L.; Liu,~Y.; Furst,~E.~M. Directed Colloidal Self-Assembly in Toggled Magnetic Fields. \emph{Soft Matter} \textbf{2014}, \emph{10}, 1102--1109\relax
\mciteBstWouldAddEndPuncttrue
\mciteSetBstMidEndSepPunct{\mcitedefaultmidpunct}
{\mcitedefaultendpunct}{\mcitedefaultseppunct}\relax
\EndOfBibitem
\bibitem[Bauer \latin{et~al.}(2015)Bauer, Liu, Kurian, Swan, and Furst]{bauer2015}
Bauer,~J.~L.; Liu,~Y.; Kurian,~M.~J.; Swan,~J.~W.; Furst,~E.~M. Coarsening Mechanics of a Colloidal Suspension in Toggled Fields. \emph{The Journal of Chemical Physics} \textbf{2015}, \emph{143}, 074901\relax
\mciteBstWouldAddEndPuncttrue
\mciteSetBstMidEndSepPunct{\mcitedefaultmidpunct}
{\mcitedefaultendpunct}{\mcitedefaultseppunct}\relax
\EndOfBibitem
\bibitem[Kim \latin{et~al.}(2020)Kim, Sau, and Furst]{kim2020}
Kim,~H.; Sau,~M.; Furst,~E.~M. An {{Expanded State Diagram}} for the {{Directed Self-Assembly}} of {{Colloidal Suspensions}} in {{Toggled Fields}}. \emph{Langmuir} \textbf{2020}, \emph{36}, 9926--9934\relax
\mciteBstWouldAddEndPuncttrue
\mciteSetBstMidEndSepPunct{\mcitedefaultmidpunct}
{\mcitedefaultendpunct}{\mcitedefaultseppunct}\relax
\EndOfBibitem
\bibitem[{Garc{\'i}a-Ruiz}(2023)]{garcia-ruiz2023}
{Garc{\'i}a-Ruiz},~J.~M. A Fluctuating Solution to the Dolomite Problem. \emph{Science} \textbf{2023}, \emph{382}, 883--884\relax
\mciteBstWouldAddEndPuncttrue
\mciteSetBstMidEndSepPunct{\mcitedefaultmidpunct}
{\mcitedefaultendpunct}{\mcitedefaultseppunct}\relax
\EndOfBibitem
\bibitem[Kim \latin{et~al.}(2023)Kim, Kimura, Puchala, Yamazaki, Becker, and Sun]{kim2023}
Kim,~J.; Kimura,~Y.; Puchala,~B.; Yamazaki,~T.; Becker,~U.; Sun,~W. Dissolution Enables Dolomite Crystal Growth near Ambient Conditions. \emph{Science} \textbf{2023}, \emph{382}, 915--920\relax
\mciteBstWouldAddEndPuncttrue
\mciteSetBstMidEndSepPunct{\mcitedefaultmidpunct}
{\mcitedefaultendpunct}{\mcitedefaultseppunct}\relax
\EndOfBibitem
\bibitem[Klotsa and Jack(2013)Klotsa, and Jack]{klotsa2013}
Klotsa,~D.; Jack,~R.~L. Controlling Crystal Self-Assembly Using a Real-Time Feedback Scheme. \emph{The Journal of Chemical Physics} \textbf{2013}, \emph{138}, 094502\relax
\mciteBstWouldAddEndPuncttrue
\mciteSetBstMidEndSepPunct{\mcitedefaultmidpunct}
{\mcitedefaultendpunct}{\mcitedefaultseppunct}\relax
\EndOfBibitem
\bibitem[Chennakesavalu and Rotskoff(2021)Chennakesavalu, and Rotskoff]{chennakesavalu2021}
Chennakesavalu,~S.; Rotskoff,~G.~M. Probing the Theoretical and Computational Limits of Dissipative Design. \emph{The Journal of Chemical Physics} \textbf{2021}, \emph{155}, 194114\relax
\mciteBstWouldAddEndPuncttrue
\mciteSetBstMidEndSepPunct{\mcitedefaultmidpunct}
{\mcitedefaultendpunct}{\mcitedefaultseppunct}\relax
\EndOfBibitem
\bibitem[Ju{\'a}rez and Bevan(2012)Ju{\'a}rez, and Bevan]{juarez2012}
Ju{\'a}rez,~J.~J.; Bevan,~M.~A. Feedback {{Controlled Colloidal Self-Assembly}}. \emph{Advanced Functional Materials} \textbf{2012}, \emph{22}, 3833--3839\relax
\mciteBstWouldAddEndPuncttrue
\mciteSetBstMidEndSepPunct{\mcitedefaultmidpunct}
{\mcitedefaultendpunct}{\mcitedefaultseppunct}\relax
\EndOfBibitem
\bibitem[Xue \latin{et~al.}(2014)Xue, {Beltran-Villegas}, Tang, Bevan, and Grover]{xue2014}
Xue,~Y.; {Beltran-Villegas},~D.~J.; Tang,~X.; Bevan,~M.~A.; Grover,~M.~A. Optimal {{Design}} of a {{Colloidal Self-Assembly Process}}. \emph{IEEE Transactions on Control Systems Technology} \textbf{2014}, \emph{22}, 1956--1963\relax
\mciteBstWouldAddEndPuncttrue
\mciteSetBstMidEndSepPunct{\mcitedefaultmidpunct}
{\mcitedefaultendpunct}{\mcitedefaultseppunct}\relax
\EndOfBibitem
\bibitem[Tang \latin{et~al.}(2016)Tang, Rupp, Yang, Edwards, Grover, and Bevan]{tang2016a}
Tang,~X.; Rupp,~B.; Yang,~Y.; Edwards,~T.~D.; Grover,~M.~A.; Bevan,~M.~A. Optimal {{Feedback Controlled Assembly}} of {{Perfect Crystals}}. \emph{ACS Nano} \textbf{2016}, \emph{10}, 6791--6798\relax
\mciteBstWouldAddEndPuncttrue
\mciteSetBstMidEndSepPunct{\mcitedefaultmidpunct}
{\mcitedefaultendpunct}{\mcitedefaultseppunct}\relax
\EndOfBibitem
\bibitem[Whitelam and Tamblyn(2020)Whitelam, and Tamblyn]{whitelam2020}
Whitelam,~S.; Tamblyn,~I. Learning to Grow: {{Control}} of Material Self-Assembly Using Evolutionary Reinforcement Learning. \emph{Physical Review E} \textbf{2020}, \emph{101}, 052604\relax
\mciteBstWouldAddEndPuncttrue
\mciteSetBstMidEndSepPunct{\mcitedefaultmidpunct}
{\mcitedefaultendpunct}{\mcitedefaultseppunct}\relax
\EndOfBibitem
\bibitem[Trubiano and Hagan(2022)Trubiano, and Hagan]{trubiano2022}
Trubiano,~A.; Hagan,~M.~F. Optimization of Non-Equilibrium Self-Assembly Protocols Using {{Markov}} State Models. \emph{The Journal of Chemical Physics} \textbf{2022}, \emph{157}, 244901\relax
\mciteBstWouldAddEndPuncttrue
\mciteSetBstMidEndSepPunct{\mcitedefaultmidpunct}
{\mcitedefaultendpunct}{\mcitedefaultseppunct}\relax
\EndOfBibitem
\bibitem[{Mart{\'i}n-Roca} \latin{et~al.}(2023){Mart{\'i}n-Roca}, {Horcajo-Fern{\'a}ndez}, Valeriani, G{\'a}mez, and {Mart{\'i}nez-Pedrero}]{martin-roca2023}
{Mart{\'i}n-Roca},~J.; {Horcajo-Fern{\'a}ndez},~M.; Valeriani,~C.; G{\'a}mez,~F.; {Mart{\'i}nez-Pedrero},~F. Field-{{Pulse-Induced Annealing}} of {{2D Colloidal Polycrystals}}. \emph{Nanomaterials} \textbf{2023}, \emph{13}, 397\relax
\mciteBstWouldAddEndPuncttrue
\mciteSetBstMidEndSepPunct{\mcitedefaultmidpunct}
{\mcitedefaultendpunct}{\mcitedefaultseppunct}\relax
\EndOfBibitem
\bibitem[Jha \latin{et~al.}(2012)Jha, Kuzovkov, Grzybowski, and {Olvera de la Cruz}]{jha2012}
Jha,~P.~K.; Kuzovkov,~V.; Grzybowski,~B.~A.; {Olvera de la Cruz},~M. Dynamic Self-Assembly of Photo-Switchable Nanoparticles. \emph{Soft Matter} \textbf{2012}, \emph{8}, 227--234\relax
\mciteBstWouldAddEndPuncttrue
\mciteSetBstMidEndSepPunct{\mcitedefaultmidpunct}
{\mcitedefaultendpunct}{\mcitedefaultseppunct}\relax
\EndOfBibitem
\bibitem[Risbud and Swan(2015)Risbud, and Swan]{risbud2015}
Risbud,~S.~R.; Swan,~J.~W. Dynamic Self-Assembly of Colloids through Periodic Variation of Inter-Particle Potentials. \emph{Soft Matter} \textbf{2015}, \emph{11}, 3232--3240\relax
\mciteBstWouldAddEndPuncttrue
\mciteSetBstMidEndSepPunct{\mcitedefaultmidpunct}
{\mcitedefaultendpunct}{\mcitedefaultseppunct}\relax
\EndOfBibitem
\bibitem[Kao \latin{et~al.}(2021)Kao, VanSaders, Glotzer, and Solomon]{kao2021}
Kao,~P.-K.; VanSaders,~B.~J.; Glotzer,~S.~C.; Solomon,~M.~J. Accelerated Annealing of Colloidal Crystal Monolayers by Means of Cyclically Applied Electric Fields. \emph{Scientific Reports} \textbf{2021}, \emph{11}, 11042\relax
\mciteBstWouldAddEndPuncttrue
\mciteSetBstMidEndSepPunct{\mcitedefaultmidpunct}
{\mcitedefaultendpunct}{\mcitedefaultseppunct}\relax
\EndOfBibitem
\bibitem[Chen \latin{et~al.}(2022)Chen, Sun, Zhu, Li, and Sun]{chen2022a}
Chen,~Z.-Q.; Sun,~Y.-W.; Zhu,~Y.-L.; Li,~Z.-W.; Sun,~Z.-Y. A Chiral Smectic Phase Induced by an Alternating External Field. \emph{Soft Matter} \textbf{2022}, \emph{18}, 2569--2576\relax
\mciteBstWouldAddEndPuncttrue
\mciteSetBstMidEndSepPunct{\mcitedefaultmidpunct}
{\mcitedefaultendpunct}{\mcitedefaultseppunct}\relax
\EndOfBibitem
\bibitem[Pigard and M{\"u}ller(2019)Pigard, and M{\"u}ller]{pigard2019a}
Pigard,~L.; M{\"u}ller,~M. Interface {{Repulsion}} and {{Lamellar Structures}} in {{Thin Films}} of {{Homopolymer Blends}} Due to {{Thermal Oscillations}}. \emph{Physical Review Letters} \textbf{2019}, \emph{122}, 237801\relax
\mciteBstWouldAddEndPuncttrue
\mciteSetBstMidEndSepPunct{\mcitedefaultmidpunct}
{\mcitedefaultendpunct}{\mcitedefaultseppunct}\relax
\EndOfBibitem
\bibitem[Weeks \latin{et~al.}(1971)Weeks, Chandler, and Andersen]{weeks1971}
Weeks,~J.~D.; Chandler,~D.; Andersen,~H.~C. Role of {{Repulsive Forces}} in {{Determining}} the {{Equilibrium Structure}} of {{Simple Liquids}}. \emph{The Journal of Chemical Physics} \textbf{1971}, \emph{54}, 5237--5247\relax
\mciteBstWouldAddEndPuncttrue
\mciteSetBstMidEndSepPunct{\mcitedefaultmidpunct}
{\mcitedefaultendpunct}{\mcitedefaultseppunct}\relax
\EndOfBibitem
\bibitem[Jack \latin{et~al.}(2007)Jack, Hagan, and Chandler]{jack2007}
Jack,~R.~L.; Hagan,~M.~F.; Chandler,~D. Fluctuation-Dissipation Ratios in the Dynamics of Self-Assembly. \emph{Physical Review E} \textbf{2007}, \emph{76}, 021119\relax
\mciteBstWouldAddEndPuncttrue
\mciteSetBstMidEndSepPunct{\mcitedefaultmidpunct}
{\mcitedefaultendpunct}{\mcitedefaultseppunct}\relax
\EndOfBibitem
\bibitem[Nguyen \latin{et~al.}(2007)Nguyen, Reddy, and Brooks]{nguyen2007}
Nguyen,~H.~D.; Reddy,~V.~S.; Brooks,~C.~L. Deciphering the {{Kinetic Mechanism}} of {{Spontaneous Self-Assembly}} of {{Icosahedral Capsids}}. \emph{Nano Letters} \textbf{2007}, \emph{7}, 338--344\relax
\mciteBstWouldAddEndPuncttrue
\mciteSetBstMidEndSepPunct{\mcitedefaultmidpunct}
{\mcitedefaultendpunct}{\mcitedefaultseppunct}\relax
\EndOfBibitem
\bibitem[Rapaport(2004)]{rapaport2004}
Rapaport,~D.~C. Self-Assembly of Polyhedral Shells: {{A}} Molecular Dynamics Study. \emph{Physical Review E} \textbf{2004}, \emph{70}, 051905\relax
\mciteBstWouldAddEndPuncttrue
\mciteSetBstMidEndSepPunct{\mcitedefaultmidpunct}
{\mcitedefaultendpunct}{\mcitedefaultseppunct}\relax
\EndOfBibitem
\bibitem[Rapaport(2008)]{rapaport2008}
Rapaport,~D.~C. Role of {{Reversibility}} in {{Viral Capsid Growth}}: {{A Paradigm}} for {{Self-Assembly}}. \emph{Physical Review Letters} \textbf{2008}, \emph{101}, 186101\relax
\mciteBstWouldAddEndPuncttrue
\mciteSetBstMidEndSepPunct{\mcitedefaultmidpunct}
{\mcitedefaultendpunct}{\mcitedefaultseppunct}\relax
\EndOfBibitem
\bibitem[Krishna \latin{et~al.}(2010)Krishna, Ayton, and Voth]{krishna2010a}
Krishna,~V.; Ayton,~G.~S.; Voth,~G.~A. Role of {{Protein Interactions}} in {{Defining HIV-1 Viral Capsid Shape}} and {{Stability}}: {{A Coarse-Grained Analysis}}. \emph{Biophysical Journal} \textbf{2010}, \emph{98}, 18--26\relax
\mciteBstWouldAddEndPuncttrue
\mciteSetBstMidEndSepPunct{\mcitedefaultmidpunct}
{\mcitedefaultendpunct}{\mcitedefaultseppunct}\relax
\EndOfBibitem
\bibitem[Perlmutter and Hagan(2015)Perlmutter, and Hagan]{perlmutter2015}
Perlmutter,~J.~D.; Hagan,~M.~F. Mechanisms of {{Virus Assembly}}. \emph{Annual Review of Physical Chemistry} \textbf{2015}, \emph{66}, 217--239\relax
\mciteBstWouldAddEndPuncttrue
\mciteSetBstMidEndSepPunct{\mcitedefaultmidpunct}
{\mcitedefaultendpunct}{\mcitedefaultseppunct}\relax
\EndOfBibitem
\bibitem[Pak \latin{et~al.}(2019)Pak, Grime, Yu, and Voth]{pak2019}
Pak,~A.~J.; Grime,~J. M.~A.; Yu,~A.; Voth,~G.~A. Off-{{Pathway Assembly}}: {{A Broad-Spectrum Mechanism}} of {{Action}} for {{Drugs That Undermine Controlled HIV-1 Viral Capsid Formation}}. \emph{Journal of the American Chemical Society} \textbf{2019}, \emph{141}, 10214--10224\relax
\mciteBstWouldAddEndPuncttrue
\mciteSetBstMidEndSepPunct{\mcitedefaultmidpunct}
{\mcitedefaultendpunct}{\mcitedefaultseppunct}\relax
\EndOfBibitem
\bibitem[Hagan and Grason(2021)Hagan, and Grason]{hagan2021}
Hagan,~M.~F.; Grason,~G.~M. Equilibrium Mechanisms of Self-Limiting Assembly. \emph{Reviews of Modern Physics} \textbf{2021}, \emph{93}, 025008\relax
\mciteBstWouldAddEndPuncttrue
\mciteSetBstMidEndSepPunct{\mcitedefaultmidpunct}
{\mcitedefaultendpunct}{\mcitedefaultseppunct}\relax
\EndOfBibitem
\bibitem[Kashchiev \latin{et~al.}(2005)Kashchiev, Vekilov, and Kolomeisky]{kashchiev2005}
Kashchiev,~D.; Vekilov,~P.~G.; Kolomeisky,~A.~B. Kinetics of Two-Step Nucleation of Crystals. \emph{The Journal of Chemical Physics} \textbf{2005}, \emph{122}, 244706\relax
\mciteBstWouldAddEndPuncttrue
\mciteSetBstMidEndSepPunct{\mcitedefaultmidpunct}
{\mcitedefaultendpunct}{\mcitedefaultseppunct}\relax
\EndOfBibitem
\bibitem[De~Yoreo \latin{et~al.}(2015)De~Yoreo, Gilbert, Sommerdijk, Penn, Whitelam, Joester, Zhang, Rimer, Navrotsky, Banfield, Wallace, Michel, Meldrum, C{\"o}lfen, and Dove]{deyoreo2015}
De~Yoreo,~J.~J.; Gilbert,~P. U. P.~A.; Sommerdijk,~N. A. J.~M.; Penn,~R.~L.; Whitelam,~S.; Joester,~D.; Zhang,~H.; Rimer,~J.~D.; Navrotsky,~A.; Banfield,~J.~F.; Wallace,~A.~F.; Michel,~F.~M.; Meldrum,~F.~C.; C{\"o}lfen,~H.; Dove,~P.~M. Crystallization by Particle Attachment in Synthetic, Biogenic, and Geologic Environments. \emph{Science} \textbf{2015}, \emph{349}, aaa6760\relax
\mciteBstWouldAddEndPuncttrue
\mciteSetBstMidEndSepPunct{\mcitedefaultmidpunct}
{\mcitedefaultendpunct}{\mcitedefaultseppunct}\relax
\EndOfBibitem
\bibitem[Dogan \latin{et~al.}(2014)Dogan, Gianni, and Jemth]{dogan2014}
Dogan,~J.; Gianni,~S.; Jemth,~P. The Binding Mechanisms of Intrinsically Disordered Proteins. \emph{Phys. Chem. Chem. Phys.} \textbf{2014}, \emph{16}, 6323--6331\relax
\mciteBstWouldAddEndPuncttrue
\mciteSetBstMidEndSepPunct{\mcitedefaultmidpunct}
{\mcitedefaultendpunct}{\mcitedefaultseppunct}\relax
\EndOfBibitem
\bibitem[Wilson \latin{et~al.}(2018)Wilson, Bommarius, Champion, Chernoff, Lynn, Paravastu, Liang, Hsieh, and Heemstra]{wilson2018}
Wilson,~C.~J.; Bommarius,~A.~S.; Champion,~J.~A.; Chernoff,~Y.~O.; Lynn,~D.~G.; Paravastu,~A.~K.; Liang,~C.; Hsieh,~M.-C.; Heemstra,~J.~M. Biomolecular {{Assemblies}}: {{Moving}} from {{Observation}} to {{Predictive Design}}. \emph{Chemical Reviews} \textbf{2018}, \emph{118}, 11519--11574\relax
\mciteBstWouldAddEndPuncttrue
\mciteSetBstMidEndSepPunct{\mcitedefaultmidpunct}
{\mcitedefaultendpunct}{\mcitedefaultseppunct}\relax
\EndOfBibitem
\bibitem[Sherman and Swan(2016)Sherman, and Swan]{sherman2016}
Sherman,~Z.~M.; Swan,~J.~W. Dynamic, {{Directed Self-Assembly}} of {{Nanoparticles}} {\emph{via}} {{Toggled Interactions}}. \emph{ACS Nano} \textbf{2016}, \emph{10}, 5260--5271\relax
\mciteBstWouldAddEndPuncttrue
\mciteSetBstMidEndSepPunct{\mcitedefaultmidpunct}
{\mcitedefaultendpunct}{\mcitedefaultseppunct}\relax
\EndOfBibitem
\bibitem[Sherman and Swan(2019)Sherman, and Swan]{sherman2019}
Sherman,~Z.~M.; Swan,~J.~W. Transmutable {{Colloidal Crystals}} and {{Active Phase Separation}} {\emph{via}} {{Dynamic}}, {{Directed Self-Assembly}} with {{Toggled External Fields}}. \emph{ACS Nano} \textbf{2019}, \emph{13}, 764--771\relax
\mciteBstWouldAddEndPuncttrue
\mciteSetBstMidEndSepPunct{\mcitedefaultmidpunct}
{\mcitedefaultendpunct}{\mcitedefaultseppunct}\relax
\EndOfBibitem
\bibitem[Savage and Dinsmore(2009)Savage, and Dinsmore]{savage2009}
Savage,~J.~R.; Dinsmore,~A.~D. Experimental {{Evidence}} for {{Two-Step Nucleation}} in {{Colloidal Crystallization}}. \emph{Physical Review Letters} \textbf{2009}, \emph{102}, 198302\relax
\mciteBstWouldAddEndPuncttrue
\mciteSetBstMidEndSepPunct{\mcitedefaultmidpunct}
{\mcitedefaultendpunct}{\mcitedefaultseppunct}\relax
\EndOfBibitem
\bibitem[Sherman \latin{et~al.}(2018)Sherman, Rosenthal, and Swan]{sherman2018}
Sherman,~Z.~M.; Rosenthal,~H.; Swan,~J.~W. Phase {{Separation Kinetics}} of {{Dynamically Self-Assembling Nanoparticles}} with {{Toggled Interactions}}. \emph{Langmuir} \textbf{2018}, \emph{34}, 1029--1041\relax
\mciteBstWouldAddEndPuncttrue
\mciteSetBstMidEndSepPunct{\mcitedefaultmidpunct}
{\mcitedefaultendpunct}{\mcitedefaultseppunct}\relax
\EndOfBibitem
\end{mcitethebibliography}


\providecommand{\latin}[1]{#1}
\makeatletter
\providecommand{\doi}
  {\begingroup\let\do\@makeother\dospecials
  \catcode`\{=1 \catcode`\}=2 \doi@aux}
\providecommand{\doi@aux}[1]{\endgroup\texttt{#1}}
\makeatother
\providecommand*\mcitethebibliography{\thebibliography}
\csname @ifundefined\endcsname{endmcitethebibliography}  {\let\endmcitethebibliography\endthebibliography}{}
\begin{mcitethebibliography}{5}
\providecommand*\natexlab[1]{#1}
\providecommand*\mciteSetBstSublistMode[1]{}
\providecommand*\mciteSetBstMaxWidthForm[2]{}
\providecommand*\mciteBstWouldAddEndPuncttrue
  {\def\EndOfBibitem{\unskip.}}
\providecommand*\mciteBstWouldAddEndPunctfalse
  {\let\EndOfBibitem\relax}
\providecommand*\mciteSetBstMidEndSepPunct[3]{}
\providecommand*\mciteSetBstSublistLabelBeginEnd[3]{}
\providecommand*\EndOfBibitem{}
\mciteSetBstSublistMode{f}
\mciteSetBstMaxWidthForm{subitem}{(\alph{mcitesubitemcount})}
\mciteSetBstSublistLabelBeginEnd
  {\mcitemaxwidthsubitemform\space}
  {\relax}
  {\relax}

\bibitem[Weeks \latin{et~al.}(1971)Weeks, Chandler, and Andersen]{weeks1971}
Weeks,~J.~D.; Chandler,~D.; Andersen,~H.~C. Role of {{Repulsive Forces}} in {{Determining}} the {{Equilibrium Structure}} of {{Simple Liquids}}. \emph{The Journal of Chemical Physics} \textbf{1971}, \emph{54}, 5237--5247, DOI: \doi{10.1063/1.1674820}\relax
\mciteBstWouldAddEndPuncttrue
\mciteSetBstMidEndSepPunct{\mcitedefaultmidpunct}
{\mcitedefaultendpunct}{\mcitedefaultseppunct}\relax
\EndOfBibitem
\bibitem[Plimpton(1995)]{plimpton1995}
Plimpton,~S. Fast {{Parallel Algorithms}} for {{Short}}{\textendash}{{Range Molecular Dynamics}}. \emph{Journal of Computational Physics} \textbf{1995}, \emph{117}, 42, DOI: \doi{10.1006/jcph.1995.1039}\relax
\mciteBstWouldAddEndPuncttrue
\mciteSetBstMidEndSepPunct{\mcitedefaultmidpunct}
{\mcitedefaultendpunct}{\mcitedefaultseppunct}\relax
\EndOfBibitem
\bibitem[Tagliazucchi \latin{et~al.}(2014)Tagliazucchi, Weiss, and Szleifer]{tagliazucchi2014}
Tagliazucchi,~M.; Weiss,~E.~A.; Szleifer,~I. Dissipative Self-Assembly of Particles Interacting through Time-Oscillatory Potentials. \emph{Proceedings of the National Academy of Sciences} \textbf{2014}, \emph{111}, 9751--9756, DOI: \doi{10.1073/pnas.1406122111}\relax
\mciteBstWouldAddEndPuncttrue
\mciteSetBstMidEndSepPunct{\mcitedefaultmidpunct}
{\mcitedefaultendpunct}{\mcitedefaultseppunct}\relax
\EndOfBibitem
\bibitem[Tagliazucchi and Szleifer(2016)Tagliazucchi, and Szleifer]{tagliazucchi2016}
Tagliazucchi,~M.; Szleifer,~I. Dynamics of Dissipative Self-Assembly of Particles Interacting through Oscillatory Forces. \emph{Faraday Discussions} \textbf{2016}, \emph{186}, 399--418, DOI: \doi{10.1039/C5FD00115C}\relax
\mciteBstWouldAddEndPuncttrue
\mciteSetBstMidEndSepPunct{\mcitedefaultmidpunct}
{\mcitedefaultendpunct}{\mcitedefaultseppunct}\relax
\EndOfBibitem
\end{mcitethebibliography}




\clearpage

\makeatletter\@input{xx.tex}\makeatother

\end{document}


\maketitle

\section{Additional Methods Details}\label{sec:moreMethods}

\subsection{Particle and Inter-particle Interactions Details}

The triangular particles are composed of fifteen, partially overlapping circular subparticles that are rigidly held together. Each triangular edge is composed of six subparticles, the outermost of which serve as the vertices of the triangular particle and are shared with the neighboring edge. The triangular particles have an edge length of 1$\sigma$, while the subparticles have a diameter of $\sigma_{\mathrm{LJ}} = 0.25\sigma$, where $\sigma$ is the lengthscale of the system in reduced units.

The subparticles on two edges of the equilateral triangle are assigned type A, while the subparticles on the third edge (including their vertices) are assigned type B. The type A - type A subparticles of differing triangular particles interact via a Lennard-Jones (LJ) potential, which is illustrated in Figure \ref{fgr:structures}a, and defined as:
\begin{equation}
    V_\mathrm{LJ}(r_{ij}) = 4\epsilon_{ij}[(\frac{\sigma_{\mathrm{LJ}}}{r_{ij}})^{12}-(\frac{\sigma_{\mathrm{LJ}}}{r_{ij}})^{6}] + c \quad \quad r_{ij} < r_{\mathrm{c}},
    \label{eq:ljPotential}
\end{equation}
where $r_{ij}$ is the distance between the centers of an $ij$-pair, $\sigma_{\mathrm{ LJ}}$ is the distance in which the potential goes to zero, and $\epsilon$ is the depth of the potential well. For static simulations, $\epsilon$ is held constant throughout the simulation, while $\epsilon$ is oscillated in time during oscillatory simulations as detailed in the main text. $c$ is the value necessary to shift the potential so that $V_\mathrm{LJ}(r_\mathrm{c})$ = 0. 
For type A subparticles, $\sigma_{ \mathrm{LJ}}$ defines the spatial extent of the LJ interaction, and the cutoff distance is set so that $r_\mathrm{c} = 2.5\sigma_{ \mathrm{LJ}}$.  

The type B - type B  and type A - type B
subparticles interact via the Weeks-Chandler-Andersen potential,\cite{weeks1971} which truncates and shifts 
the LJ potential given in Equation~\ref{eq:ljPotential} to only include the repulsive portion of the curve where $\frac{\partial U_{ij} (r)}{\partial r} < 0$. 
In order to truncate and shift the potential,
we set $\epsilon_{AB} = \epsilon_{BB}$ = 1.0 and 
$r_\mathrm{c}$
=$2^{1/6}\sigma_{\mathrm{LJ}}$.

\subsection{Simulation Details}

All simulations are performed using Langevin dynamics, as implemented in LAMMPS.\cite{plimpton1995} Simulations started with a randomly distributed system of 150 triangular particles in a periodic box of size $L$ x $L$ with no particle overlap. Simulations with a volume fraction of $\phi$ = 0.1 were done in a box with dimensions 25.48$\sigma$ x 25.48$\sigma$, while simulations at a volume fraction of $\phi$ = 0.005 occurred in a box with dimensions of 113.98$\sigma$ x 113.98$\sigma$.

To relax the system, initial velocities were assigned and the simulation was allowed to equilibrate for 1,000,000 steps (5,000$\tau$). During this process, all subparticles interacted via the WCA potential, as this ensures there is no particle overlap and establishes an equilibrated random distribution of triangles. After equilibration, the type A attractive interactions were turned on, and the simulation was allowed to proceed for 30,000,000 steps (150,000 $\tau$).

Simulations were evolved in time using the velocity-Verlet algorithm with a timestep of 0.005$\tau$. To mimic solvent dynamics and maintain a constant temperature, the Langevin thermostat was used. The Langevin equation is defined as
 \begin{equation}
    m \frac{dv}{dt} = -\gamma v - \frac{dU}{dr}+ \eta(t).
    \label{eq:langevin}
\end{equation}
In addition to the conservative force, $\frac{-dU}{dr}$, that arises from the inter-particle interactions, each triangular particle experiences a friction force and a random force.
 The friction force is defined as $-\gamma v$, where $\gamma$ tunes the strength of the friction and $v$ is the velocity describes the frictional interaction between the implicit solvent and the particles. In LAMMPS, we set $\gamma$ by the $damp$ parameter, with $damp = \frac{m}{\gamma}$, where $m$ is mass of the triangular body.\cite{plimpton1995} We set $damp = 0.35$ in reduced units.
 The random force, $\eta(t)$ in Equation~\ref{eq:langevin}, mimics the random bumps and kicks the solvent atoms will provide to the subparticle at temperature $T$. 
 The LAMMPS code was modified so that it used a Gaussian random number for $\eta(t)$ to ensure the appropriate fluctuation statistics.

\subsection{Data Analysis Details}

Once the attractive LJ interactions have been turned on, atom positions are recorded every 10$\tau$ to monitor the progression of the self-assembly process. To have an even sampling while calculating the final yields, the final 7,500$\tau$ of each simulation is sampled every 15$\tau$ in the static system.

In the temporally variant system, however, sampling the system with a constant interval will result in uneven sampling of the system as it switches between $\epsilon_{\mathrm{ min}}$ and $\epsilon_{\mathrm{max}}$. To ensure even sampling between the $\epsilon_{\mathrm{min}}$ and $\epsilon_{\mathrm{max}}$ half-cycles during the calculation, 
the last 7,500$\tau$ of the simulations is divided into slices of 150$\tau$ 
and ten evenly spaced snapshots from the first complete period of each slice were recorded.  
For periods equal to or shorter than 0.05$\tau$ (ten steps), 
every step of the first complete period of each slice was recorded to evenly monitor time within the $\epsilon_{\mathrm{min}}$ and $\epsilon_{\mathrm{max}}$ half-cycles. 
\subsubsection{Capsid Counting}

To determine the number of assembled capsids, the locations of the type A subparticles that are positioned at the attractive vertex of each triangular particle are compared to the attractive vertices of the other nearby triangular particles. We consider two triangular particles to be part of a potential capsid if the distance between the centers of their attractive vertices are less than 0.85$\sigma$. A distance less than 0.85$\sigma$ was determined to encompass the fluctuations possible in the capsid shape. If any subparticle tip is found to have five neighboring subparticle tips within a distance of 0.85$\sigma$, the capsid is considered assembled. 

\subsubsection{Aggregate Counting}

We also analyze the number of other aggregate structures to better understand the overall assembly process. After determining the number of capsids within the system, the particles that are not within assembled capsids are analyzed to determine what size aggregate they are assembled into. We compare the three vertices of each triangular particle with the three vertices of the nearby particles to determine if the maximum length between any two vertices is less than 2.1$\sigma$. A maximum distance of 2.1$\sigma$ ensures that two particles are assembled, but accounts for all configurations the particles can take in the snake-like aggregates. Aggregates are then built by noting which particles belong to each aggregate.

\section{Temperature in the intermediate period regime}
\begin{figure}[h]
    \centering
    \includegraphics[height=5cm]{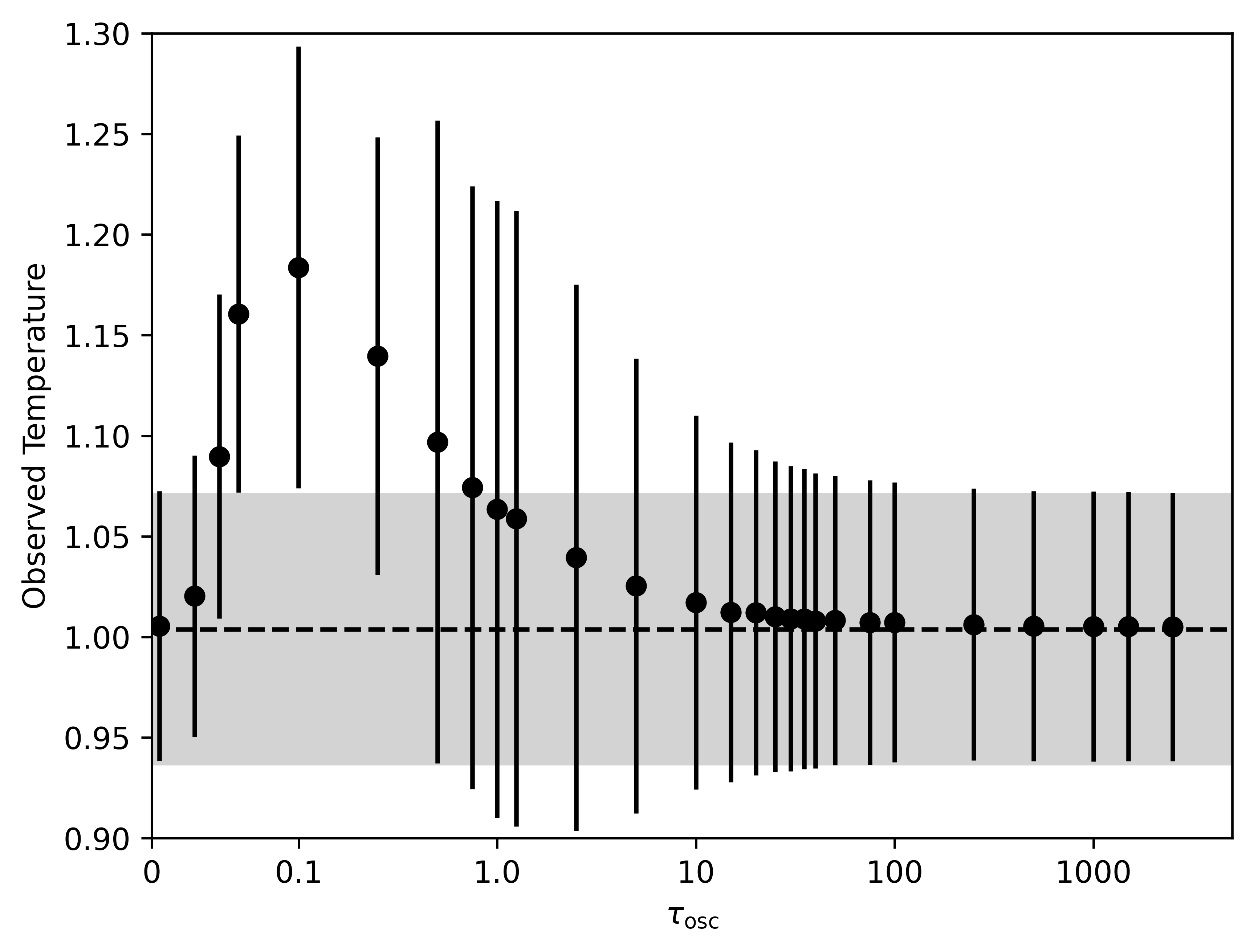}    
    \caption{{{\bf Average observed temperatures at $\epsilon = 1.25 k_{\mathrm{B}}T$.}} The average observed temperature and its standard deviation over the last 5,000$\tau$ of the simulation is plotted vs. the oscillation period $\tau_{\mathrm{osc}}$. The average observed static temperature is represented by a black dashed horizontal line, and the grey shaded region indicates its standard deviation. We present the average observed temperatures at $\epsilon = 1.25k_{\mathrm{B}}T$ for an amplitude of 0.4$k_{\mathrm{B}}T$ as this value is closest to the median $\epsilon$ value observed in Figure~\ref{img:intermediateYield}b.
    }
    \label{fig:temperatures}
    \end{figure}

Langevin Dynamics is utilized to move the simulation forward in time and maintain a constant temperature. In LAMMPS,\cite{plimpton1995} a Langevin parameter, $damp = \frac{m}{\gamma}$  
is defined in units of $\tau$ and determines how rapidly the temperature is relaxed to the target temperature\cite{plimpton1995}. 
For oscillation periods similar to the timescale of the damp parameter (0.35$\tau$), the average temperature of the simulation is higher than the target temperature of 1.0. Figure~\ref{fig:temperatures} shows the average temperatures observed in the simulations over the last 5,000$\tau$ for a range of oscillation periods as compared to the average temperature for the static potential system. Temperatures at the fast oscillation limit and at oscillations periods longer than 10$\tau$ are comparable to those in the static system, but periods from 0.04$\tau$ to 5$\tau$ deviate more than 5\% from the target temperature. 
To standardize our results for an average $\epsilon$ value over the full range of oscillation periods, we report $\epsilon_{\mathrm{avg}}$ in units of $k_{\rm B}T^{\rm(obs)}$, where $T^{\rm(obs)}$ is the average observed temperature during the run.
This rescaling was only done for Figure~\ref{img:intermediateYield}a\&b as no other figures include results from oscillation periods between 0.04$\tau$ to 5$\tau$.

\section{Kinetic traces of structure formation within the static system from $\epsilon = 0.75k_{\mathrm{B}}T$ to $\epsilon = 2.15k_{\mathrm{B}}T$.}
\begin{figure}[H]
    \centering
    \includegraphics[height=12.4cm]{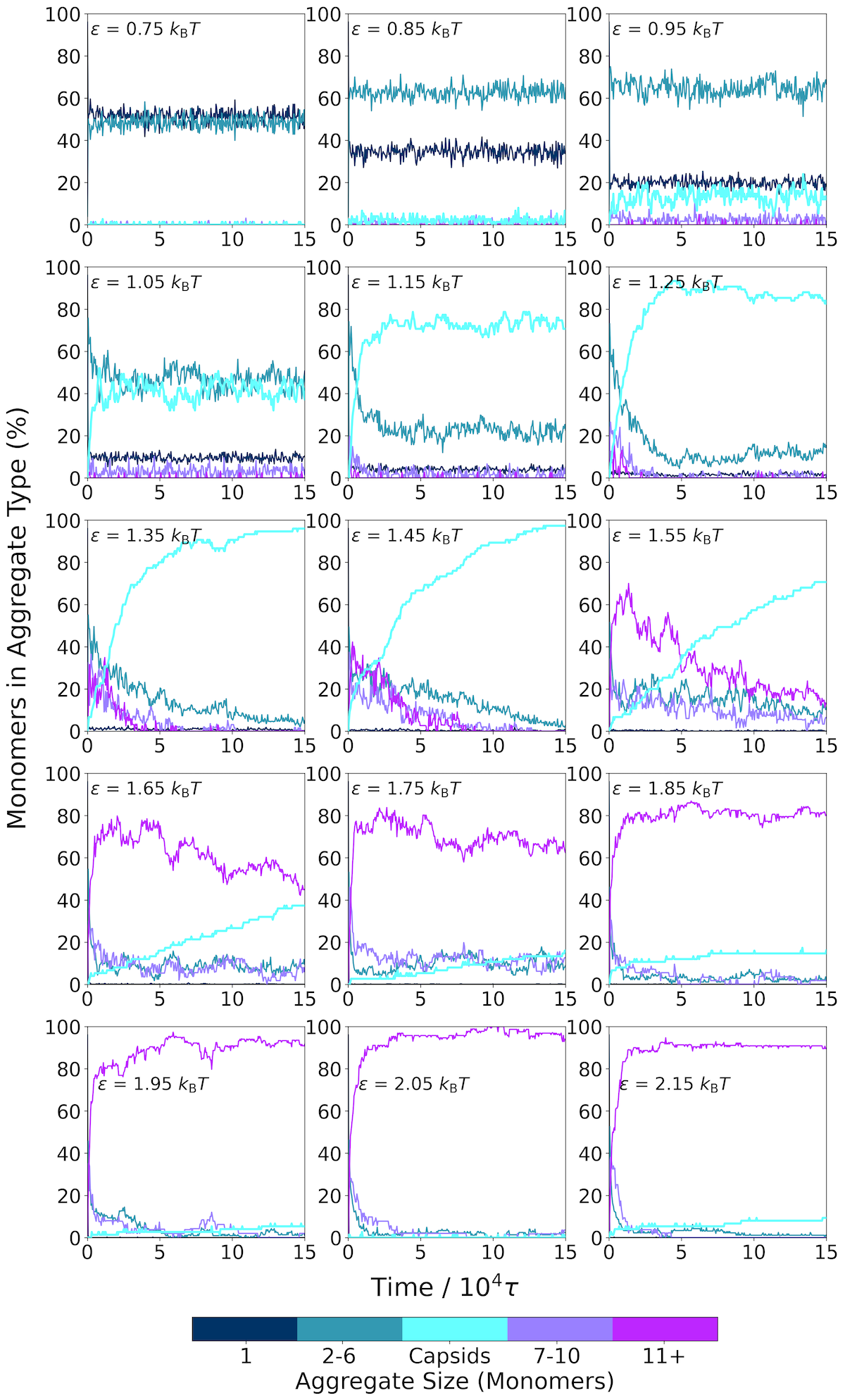}
    \caption{{\bf Formation of capsids and aggregates vs. time for different $\epsilon$ values in the non-oscillatory system.} The formation of various aggregate and capsid species is plotted here vs. time for different $\epsilon$ values over a simulation time of 150,000$\tau$. The different aggregate sizes are indicated by the color bar, and the kinetic traces are the average of three independent trajectories.}
    \label{img:EquilForm}
\end{figure}

In Figure~\ref{img:EquilForm} we expand Figure~\ref{img:equilYield}b to include additional interaction strengths.

\section{Mean square displacement and rotational relaxation of a single triangular particle.} 
\begin{figure}[h!]
\centering
\includegraphics[height=3.5cm]{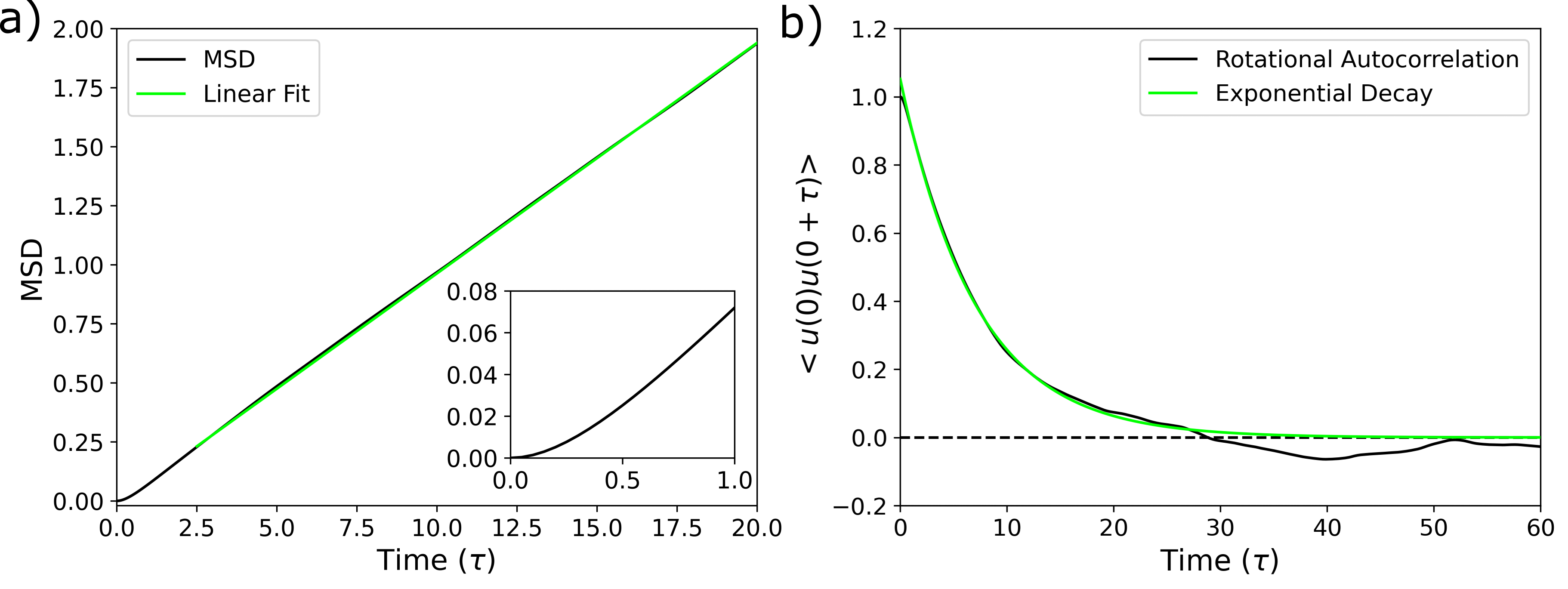}
\caption{{\bf Mean square displacement and orientational autocorrelation function for a single particle.} a)The mean square displacement is plotted vs. simulation time for a single triangular particle. Based on the linear fit line in green and the Einstein relation, the diffusion coefficient was calculated to be $D = 0.024 \pm 0.007\sigma^2/\tau$. 
b) The rotational autocorrelation function for a single triangular particle fit with $N(t) = N_0 exp(-t/\tau)$, where $\tau = 7.103\tau$, and describes the time to decay to $1/e * N_0$. The mean square displacement and rotational autocorrelation function were both calculated from 18 individual trajectories with a recording interval of 0.05$\tau$.
}
\label{img:MSD}
\end{figure}

We examine the interplay between the oscillation period and the time it takes to diffuse characteristic length-scales that characterize important energetic and structural changes. To quantify particle diffusion, we calculate the mean square displacement for a lone triangle and plot it vs. simulation time in Figure~\ref{img:MSD}a. 
Additionally, we calculate the rotational autocorrelation function and plot it against simulation time in Figure~\ref{img:MSD}b. The timescales needed to traverse characteristic distances and rotation are presented in Table~\ref{tbl:diffusiontable}.

\section{Time averaging in the fast oscillation limit.} 
Based on earlier work theoretical work by Szleifer and coworkers\cite{tagliazucchi2014,tagliazucchi2016}, the effective potential in our system can be described at the fast oscillation limit by a static LJ potential with a well depth of $\epsilon_{\mathrm{avg}}$, which is simply the time-dependent $\epsilon$ value averaged over a single oscillation period. 

To show this correspondence, we first write the effective potential at the fast oscillation limit as:
\begin{equation}
    U^{\mathrm{eff}}(r) = \int_0^{\tau_{\mathrm{osc}}} U(r,\epsilon(t))dt,
    \label{eq:effective_epsilon}
\end{equation}
where $U(r,\epsilon(t))$ is the time dependent potential energy of the system at time $t$, which is integrated over a single oscillation period, $\tau_{\mathrm{osc}}$\cite{tagliazucchi2014}. We can then substitute in the full LJ potential with a time dependent $\epsilon$ on the right hand side, such that:

\begin{equation}
    U^{\mathrm{eff}}(r) = \int_0^{\tau_{\mathrm{osc}}} 4\epsilon(t)[(\frac{\sigma}{r})^{12} - (\frac{\sigma}{r})^{6} ] dt.
    \label{eq:LJtimeO}
\end{equation}
At the fast oscillation limit, any changes to $r$ during a single oscillation period are small. As a result, Equation~\ref{eq:LJtimeO} becomes 

\begin{equation}
\begin{split}
    U^{\mathrm{eff}}(r) & = 4[(\frac{\sigma}{r})^{12} - (\frac{\sigma}{r})^{6} ] \int_0^{\tau_{\mathrm{osc}}} \epsilon(t) dt \\
    & = 4<\epsilon>_{\tau_{\mathrm{osc}}}[(\frac{\sigma}{r})^{12} - (\frac{\sigma}{r})^{6} ],
    \label{eq:LJtimeavg}
\end{split}
\end{equation}
where $<\epsilon>_{\tau_{\mathrm{osc}}}$ is the time averaged attraction strength over a single period, which we term $\epsilon_{\mathrm{avg}}$. 

Finally, since in our simulations, half the oscillation period is spent at $\epsilon_{\mathrm{max}}$ and half at $\epsilon_{\mathrm{min}}$,
\begin{equation}
\label{eq:eps_osc}
    \langle \epsilon \rangle_{\tau_{\mathrm{osc}}} = \frac{\epsilon_{\mathrm{max}} + \epsilon_{\mathrm{min}}}{2} =
    \epsilon_{\mathrm{avg}}.
\end{equation}

\section{Kinetic traces of structure formation with both static and oscillatory interactions.} 
\begin{figure}[h]
\centering
  \includegraphics[height=9.4cm]{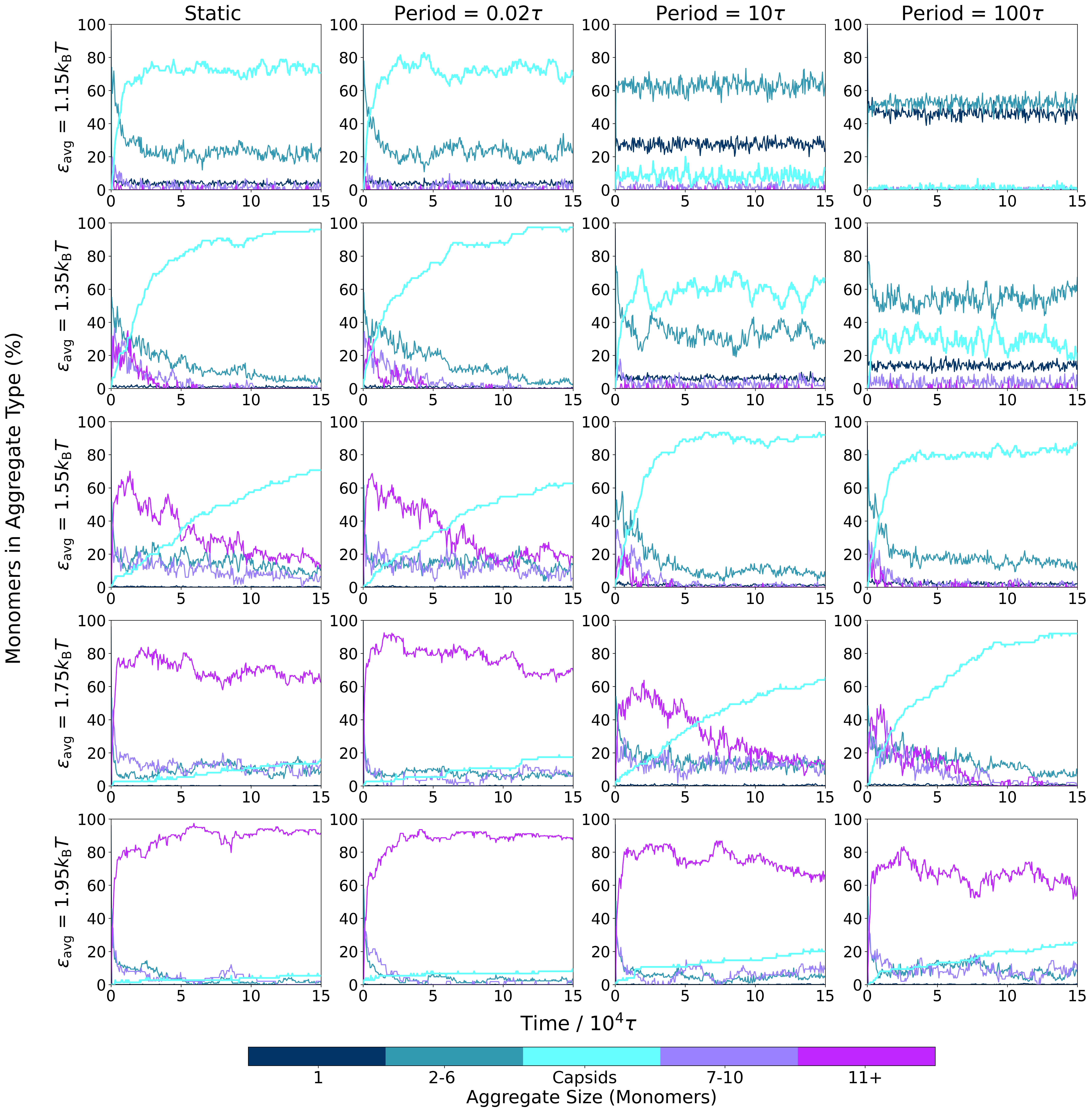}
  \caption{{\bf Formation of aggregate species over time for different oscillation periods.} The formation of various aggregate species averaged over three indepedent trajectories is shown over time for different $\epsilon$ values and oscillation periods at an amplitude of 0.4$k_{\mathrm{B}}T$. Column 1 shows results from the static potential system, Column 2 shows results from the fast oscillation limit, and Columns 3 and 4 show results from two different intermediate periods. Different assembled structures are tracked over the simulation as indicated by the color bar.
  }
  \label{fgr:formationTime}
\end{figure}

To investigate in more detail how assembly mechanisms change with oscillatory interactions, in Figure~\ref{fgr:formationTime} we compare the kinetic traces of the aggregate species in the static system (Column 1) to kinetic traces in the oscillatory system at three different periods. An oscillation period of 0.02$\tau$, which is within the fast oscillation limit, results in very similar assembly to the non-oscillatory system. Intermediate periods, shown in Columns 3 and 4, however, result in very different aggregate assembly kinetics.

\normalem
\bibliography{main}

\makeatletter\@input{xx.tex}\makeatother